\documentclass[letter,apj]{emulateapj-rtx4}
\usepackage{natbib}
\usepackage{amssymb,amsmath}
\usepackage{graphicx}

\begin{document}

\title{The Spectrum of the Diffuse Galactic Light I: The Milky Way in
  Scattered Light}

\author{Timothy D. Brandt and B. T. Draine\altaffilmark{1}}
\altaffiltext{1}{Department of Astrophysical Sciences, Peyton Hall, Princeton University, Princeton, NJ 08544, USA}

\begin{abstract}
We measure the optical spectrum of the Diffuse Galactic Light (DGL) --
the local Milky Way in reflection -- using 92,000 blank-sky spectra
from the Sloan Digital Sky Survey (SDSS).  We correlate the SDSS
optical intensity in regions of blank sky against 100 $\mu$m
intensity independently measured by the COsmic Background Explorer
(COBE) and InfraRed Astronomy Satellite (IRAS) satellites, which
provides a measure of the dust column density times the intensity of
illuminating starlight.  The spectrum of scattered light is very blue
and shows a clear 4000 \AA{} break and broad Mg b absorption.  This is
consistent with scattered starlight, and the continuum of the DGL is
well-reproduced by a simple radiative transfer model of the Galaxy.
We also detect line emission in H$\alpha$, H$\beta$, [\textsc{N\,ii}],
and [\textsc{S\,ii}], consistent with scattered light from the local
interstellar medium (ISM).  The strength of [\textsc{N\,ii}] and
[\textsc{S\,ii}], combined with upper limits on [\textsc{O\,iii}] and
[He\textsc{\,i}], indicate a relatively soft ionizing spectrum.  We
find that our measurements of the DGL can constrain dust models,
favoring a grain size distribution with relatively few large grains.
We also estimate the fraction of high-latitude H$\alpha$ which is
scattered to be $19\pm4$\%.
\end{abstract}


\section{Introduction}

All astronomical observations include light from diffuse sources other
than the target object.  In ground-based data, the dominant sources of
contamination in the optical are airglow, scattered sunlight, and
artificial sources.  Space-based missions must still contend with
zodiacal light, and all observations will include some emission and
scattering from the Galaxy's interstellar medium (ISM)
-- the diffuse Galactic light (DGL).

The first quantitative measurements of the DGL were 
photoelectric measurements by \citet{Elvey+Roach_1937}
at $\lambda\approx4500~{\rm \AA}$; after
subtracting the zodiacal and airglow contributions, the DGL was
detected for $|b|\lesssim 35^\circ$.  Subsequent studies from the ground
\citep{Elsasser+Haug_1960} and a sounding rocket \citep{Wolstencroft+Rose_1966}
found intensities at $|b|\approx 5^\circ$
corresponding to $\sim50$ 10$^{\rm th}$ magnitude stars per square
degree, or 
$\lambda I_\lambda \approx 
5\times10^{-4}~{\rm erg\, cm^{-2}\, s^{-1}\, sr}^{-1}$.
Subsequent observations from satellites
\citep{Lillie+Witt_1976,
       Morgan+Nandy+Thompson_1978,
       Henry_1981,
       Zvereva+Severnyi+Granitskii+etal_1982,
       Martin+Hurwitz+Bowyer_1990,
       Murthy+Henry+Feldman+Tennyson_1990,
       Hurwitz+Bowyer+Martin_1991,
       Murthy+Henry+Holberg_1991,
       Sasseen+Deharveng_1996,
       Seon+Edelstein+Korpela+etal_2010}
extended these studies into the vacuum ultraviolet (UV).
Most of these studies were broadband, but \citet{Martin+Hurwitz+Bowyer_1990}
detected fluorescent emission from UV-pumped H$_2$.

The spectrum of the DGL contains a wealth of information about the
physical environment where it originates and the dust that emits or
scatters it into our line-of-sight.  In this paper, we present a novel
way of measuring the spectrum of light scattered by the Galactic ISM.
This is a spectrum of the Galaxy in reflection, plus possible
luminescence from interstellar dust.

\section{Methodology} \label{sec:method}

The optical surface brightness of the DGL, $\lambda I_\lambda \approx
10^{-4}~{\rm erg\, cm^{-2}\, s^{-1}\, sr}^{-1}$, is far too low to
measure a spectrum directly.  Further, any such spectrum would be a
combination of terrestrial airglow, scattered artificial light,
zodiacal light, scattering and emission by interstellar dust, emission
by diffuse gas, and unresolved background objects.  In this section we
describe a novel technique to measure the spectrum of scattering by
the Galactic ISM.  We use 92,000 sky spectra from the Sloan Digital
Sky Survey \citep[SDSS,][]{York+Adelman+Anderson+etal_2000},
correlating their intensities against independently measured 100
$\mu$m emission to isolate the components of the DGL associated with
interstellar dust.  In a companion paper (Brandt \& Draine 2011, in
preparation)
we use the full-sky H$\alpha$ map compiled by \cite{Finkbeiner_2003}
to isolate the component of the DGL associated with emission by
diffuse \textsc{H\,ii}.

\subsection{The SDSS Sky Fibers}   \label{sec:skyfibers}

The Seventh Data Release of the SDSS
\citep{Abazajian+Adelman-McCarthy+Agueros+etal_2009} contains more
than 1.6 million spectra of stars, galaxies, and quasars, making it by
far the largest such dataset ever assembled.  The spectra were taken
by plugging 640 fibers into a $\sim$$7~\mathrm{deg}^2$ plate, with
each $2.96''$ fiber feeding the light of its target object into a pair
of spectrographs.  Each group of 640 spectra was then calibrated and
sky-subtracted by the SDSS spectroscopic pipeline
(\citealt{Stoughton+Lupton+Bernardi+etal_2002}; Burles \& Schlegel unpub.).

To obtain an accurate measurement of the sky background, a minimum of
32 fibers on each plate were placed on blank sky regions.  These
positions were relatively uniformly distributed over each plate, and
required no detection by the photometric pipeline to $5\sigma$ in any
band (Lupton, priv.~commun.).  A few ($\sim$$1$\%) of the sky fibers
were erroneously placed over bright sources, but these were flagged
and removed by the reduction pipeline, leaving a total of about 92,000
blank sky spectra used to compute the background flux.  The sky fibers
were used to construct a ``supersky'' spectrum for each plate scaled
to unit airmass.  This spectrum was then rescaled to the airmass at
each fiber (including the sky fibers themselves) and subtracted from
that fiber's spectrum.  Each plate also included 8 F dwarfs as
spectrophotometric standards, 8 F subdwarfs as reddening standards,
and 2 hot subdwarfs \citep{Stoughton+Lupton+Bernardi+etal_2002}.  The
residual sky spectra, along with all of the other spectra, were
flux-calibrated to these standards.

The sky spectra on a given plate show modest fiber-to-fiber variation,
but are dominated by noise associated with terrestrial airglow.
Hidden within this noise are real variations due to extraterrestrial
sources: zodiacal light, scattering and emission by diffuse
interstellar dust, emission by diffuse gas, and faint, unresolved
background sources.  We can isolate the components associated with
Galactic dust using the independently measured InfraRed Astronomy
Satellite (IRAS) 100 $\mu$m map, reduced and calibrated by
\cite{Schlegel+Finkbeiner+Davis_1998}, hereafter SFD.  The bulk of
this emission comes from thermally radiating dust grains heated to
$\sim$18 K by starlight.  By tracing interstellar dust illuminated by
Milky Way stars, the 100 $\mu$m map allows us to measure the spectrum
of the Galaxy in scattered light.

\subsection{Correlating Against 100 $\mu$m Intensity} \label{sec:100u}

In 1983, IRAS mapped the entire sky at 12, 25, 60, and 100 microns
\citep{Neugebauer+Habing+vanDuinen+etal_1984}.  SFD later smoothed the
100 $\mu$m map and corrected it for point sources and zodiacal light,
creating a map of diffuse Galactic infrared emission.  They further
used COsmic Background Explorer (COBE) data
\citep{Boggess+Mather+Weiss+etal_1992} at 240 $\mu$m to estimate the
temperature of the dust, ultimately producing a $\sim$6$'$ resolution
map of 100 $\mu$m emission and a lower resolution temperature map
suitable to convert 100 $\mu$m emission into a dust column density.
This map was intended mainly to estimate and correct for Galactic
extinction.  Here, we use it to correlate illuminated dust with
residual intensity in the SDSS sky fibers.

In the optically thin limit, we expect the intensity of scattered
starlight to be proportional to the column density of dust times the
intensity of the illuminating starlight.  For dust with ${\rm opacity}
\propto \nu^\beta$ near 100 $\mu$m, the intensity of the illuminating
(and hence, scattered) starlight should be proportional to $T_{\rm
dust}^{4 + \beta}$.  \cite{Planck_2011_dust} have found that the
$\lambda > 100$ $\mu$m emission is well-approximated by dust with
$\beta \approx 1.8$.  We therefore expect the SDSS sky fiber residual
intensity to be roughly proportional to $\tau_{100\,\mu \rm m}
T_{\rm dust}^{5.8}$, where the optical depth at 100 $\mu$m,
$\tau_{100\,\mu \rm m}$, is proportional to the dust column density.
We expect the 100 $\mu$m intensity itself to have a temperature
dependence.  For 18 K dust radiating at 100 $\mu$m, $h \nu / k_B T
\approx 8 \gg 1$, so that
\begin{equation}
\left( \frac{d\ln I_\nu}{d\ln T} \right) 
\bigg|_{\substack{\lambda = 100\,\mu \rm m \\ T = 18\,\rm K}} 
\approx \frac{h \nu}{k_B T} \approx 8~.
\end{equation}
We use the measured 100 $\mu$m intensity to trace the product of the
intensity of the illuminating starlight and the dust column.  In
practice, temperature variations are sufficiently small that using the
SFD column density times $T^{5.8}$ produces results indistinguishable
from those simply using 100 $\mu$m intensity.

Two complications prevent us from assuming a linear relationship
between sky fiber residuals and 100 $\mu$m intensity:
\begin{enumerate}
\item Such a model neglects self-absorption by optically thick dust.
\item The spectroscopic pipeline has already subtracted a scaled sky
  spectrum from each fiber, which includes a component of the DGL.
  This component differs from plate to plate.
\end{enumerate}
We avoid the first problem by excluding spectra and entire plates
where the dust is optically thick to visible light, with $A_V \gtrsim
0.5$ according to SFD; our results are insensitive to the precise
value of this threshold.  For a dust temperature of 18 K, this
corresponds to 100 $\mu$m emission exceeding $\sim$$10~
\mathrm{MJy}\,\mathrm{sr}^{-1}$.  We address the second point by
assuming residual optical intensity to be proportional to the excess 100
$\mu$m intensity {\it relative to the the average over that fiber's
plate}.  Our model is then
\begin{align}
\lambda I_{\lambda,\,{\rm sky},\,j,\, p} = \alpha_\lambda \left[ (\nu I_\nu)_{100\,\mu {\rm m},\,j,\, p} 
- \langle (\nu I_\nu)_{100\,\mu \rm m} \rangle_p \right]~,
\label{eq:100mumodel}
\end{align}
where $I_{\lambda,\,{\rm sky},\, j,\, p}$ is the residual intensity
in sky fiber $j$ on plate $p$ at wavelength $\lambda$,
$(I_\nu)_{100\,\mu {\rm m},\,j,\, p}$ is the 100 $\mu$m intensity
at fiber $j$'s location, $\langle\,\rangle_p$ denotes an average over
the sky fibers on plate $p$, and $\alpha_\lambda$ is a dimensionless
number that describes the relative strength of scattered and thermal
emission.  We then solve for the best-fit spectrum of coefficients
$\alpha_\lambda$, our {\it correlation spectrum}.  Defining
\begin{align}
y_{\lambda, j, p} &\equiv \lambda I_{\lambda,\,{\rm sky},\,j,\,p}
\quad {\rm and} \label{eq:ydef} \\
x_{j, p} &\equiv (\nu I_\nu)_{100\,\mu {\rm m},\,j,\,p} - \langle (\nu I_\nu)_{100\,\mu \rm m}
\rangle_p~,
\label{eq:xdef}
\end{align}
the maximum likelihood estimate for 
$\alpha_\lambda$ is
\begin{equation}
\alpha_\lambda = \left( \sum_{j, p} \frac{
\left( y_{\lambda, j, p} \right) \left( x_{j, p} \right)}
{\sigma_{\lambda, y, j, p}^2} \right) \left( \sum_{j, p}
  \frac{x_{j, p}^2}{\sigma_{\lambda, y, j, p}^2} \right)^{-1},
\label{eq:maxlike}
\end{equation}
and its variance is
\begin{equation}
\sigma^2_\lambda = \left( \sum_{j, p}
  \frac{x_{j, p}^2}{\sigma_{\lambda, y, j, p}^2} \right)^{-1}.
\label{eq:maxlikevar}
\end{equation}
In Equations \eqref{eq:maxlike} and \eqref{eq:maxlikevar},
$\sigma_{\lambda, y, j, p}^2$ is the variance of $y_{\lambda, j, p}$
as estimated by the SDSS pipeline.  

By using the residual, sky-subtracted intensity, our model adopts the
flux calibrations performed by the SDSS spectroscopic pipeline.
Beginning with the Sixth Data Release, SDSS spectra have been
flux-calibrated to PSF magnitudes appropriate for point sources rather
than fiber magnitudes appropriate for extended sources.  Dividing by
the SDSS fiber aperture therefore gives an incorrect intensity.
Figure 4 of \cite{Adelman-McCarthy+Agueros+Allam+etal_2008} shows the
difference between PSF and fiber magnitudes of point sources to be
very nearly Gaussian with a mean of 0.35 magnitudes.  The typical
seeing for spectroscopy was poor ($\sim$2$''$); as a result, the
aperture correction was found to be a very weak function of
wavelength.  We therefore recalibrate all of our spectra to fiber
magnitudes using the average flux conversion factor of $10^{0.35/2.5}
= 1.38$ before dividing by the 2.96$''$ fiber aperture.

We would like to apply the model in Equations
\eqref{eq:100mumodel}--\eqref{eq:maxlikevar} to each wavelength
observed by SDSS.  However, the wavelengths that were observed
differed slightly from night to night.  This variation is only a few
tenths of a percent, but is sufficient to blur spectral features if
not removed.  We therefore define a new wavelength array of 4000
elements, similar to but slightly larger than the 3852 element array
in most SDSS spectra.  We use cubic splines to interpolate all spectra
and errors onto this new array.  The interpolation introduces a
correlation between neighboring wavelength elements.  In Section
\ref{sec:errors}, we discuss this in detail, and demonstrate that our
measured spectra and errors are statistically very well-behaved.

\begin{figure}

\includegraphics[width=\linewidth]{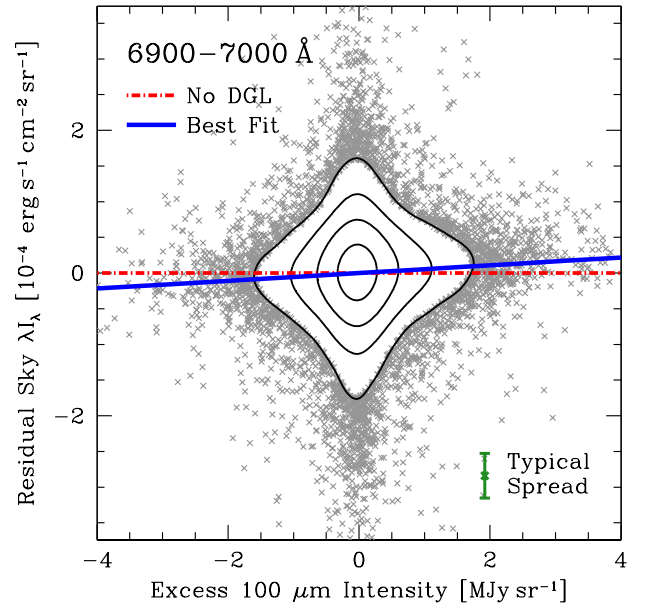}

\caption{Scatter plot showing the correlation between 100 $\mu$m
  intensity relative to the mean over an SDSS sky fiber's plate and
  the residual sky intensity averaged over the interval from 6900
  to 7000 \AA{} (Equation \ref{eq:100mumodel}).  We use
  logarithmically spaced contours where the density of points is high.
  The data demand a non-zero slope, a component of the sky background
  associated with interstellar dust, at more than $70\, \sigma$.  }

\label{fig:correlation}
\end{figure}

Figure \ref{fig:correlation} illustrates our model for the spectrum of
scattered light.  To increase the signal-to-noise ratio in Figure
\ref{fig:correlation}, we have binned each spectrum's $\sim$60
wavelength elements from 6900-7000 \AA.  Each $2.96''$ patch of blank
sky thus contributes a single point: its residual intensity averaged
over the interval from 6900 to 7000 \AA.  After discarding bad sky
fibers flagged by the spectroscopic pipeline along with fibers and
plates with $(I_\nu)_{100\,\mu \rm m} > 10$ MJy\,sr$^{-1}$
(corresponding to $A_V \gtrsim 0.5$), we are left with nearly 90,000
points representing over 5 million intensity measurements.  We
indicate these points by logarithmically spaced contours where their
density is extremely high.  We then fit them with Equation
\eqref{eq:100mumodel}, correlating the residual optical intensities
against their fibers' excess 100 $\mu$m emission.  There is a positive
correlation between the two quantities ($\alpha_\lambda > 0$), driven
by the entire ensemble of points and significant at more than
$70\,\sigma$.

The best-fit slope in Figure \ref{fig:correlation} is biased low by
its neglect of errors in the 100 $\mu$m map and by structure
unresolved by IRAS, but by a factor independent of wavelength.
Chromatic effects due to seeing and atmospheric dispersion are small
\citep{Adelman-McCarthy+Agueros+Allam+etal_2008}.  We discuss the
calibration of the correlation spectrum in Sections \ref{sec:bias} and
\ref{sec:calib}.

\subsection{Sky Coverage}
  \label{sec:skycoverage}
The SDSS focused much of its spectroscopy on the
Northern Galactic Cap because of that region's low levels of
extinction.  For the same reason, many of its plates are not very
useful for inferring the spectrum of scattered light.  The part of the
SDSS footprint that is most useful for our purposes has significant
small-scale variation in 100 $\mu$m emission, and is
disproportionately in regions of low Galactic latitude.

\begin{figure}
\includegraphics[width=\linewidth]{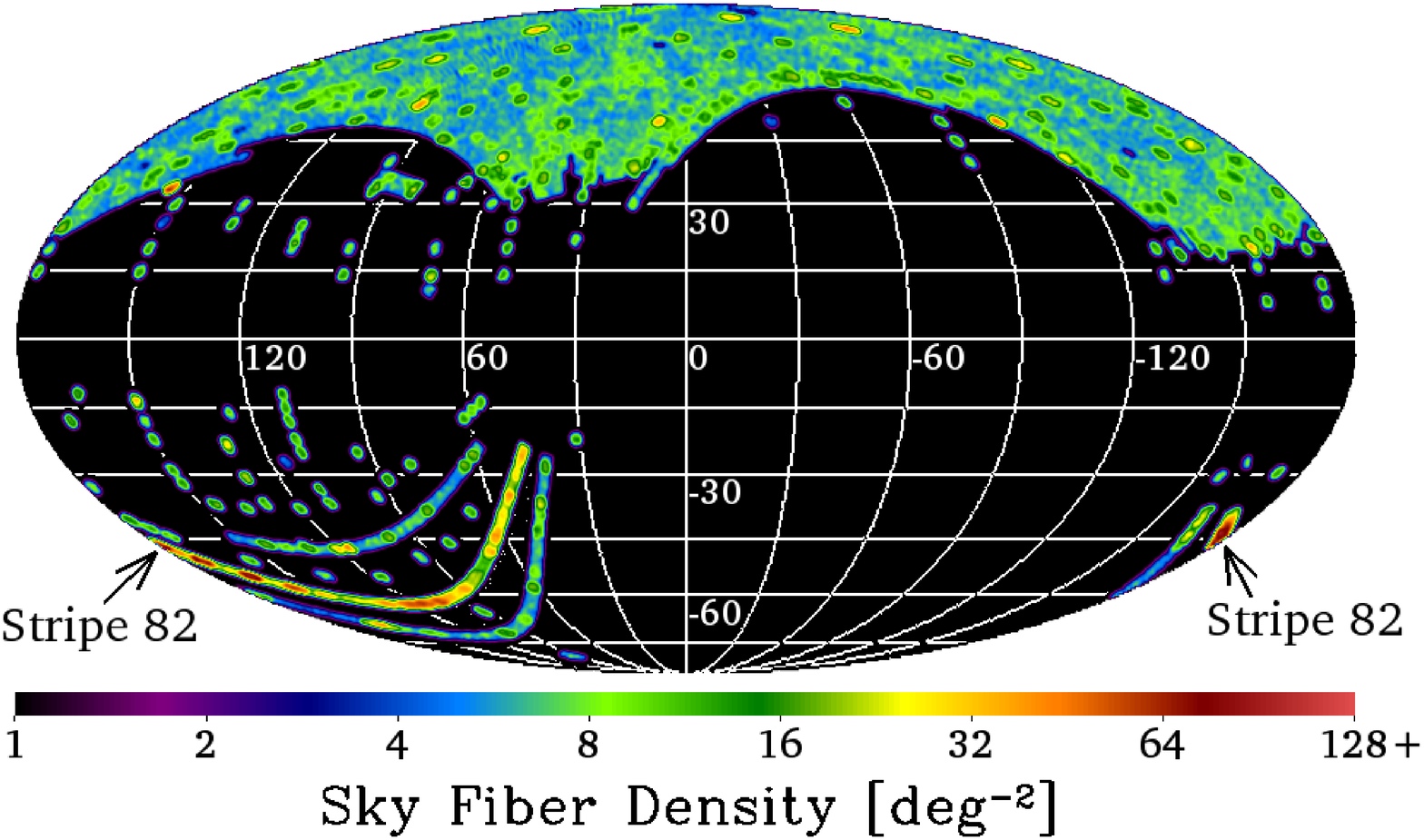}
\includegraphics[width=\linewidth]{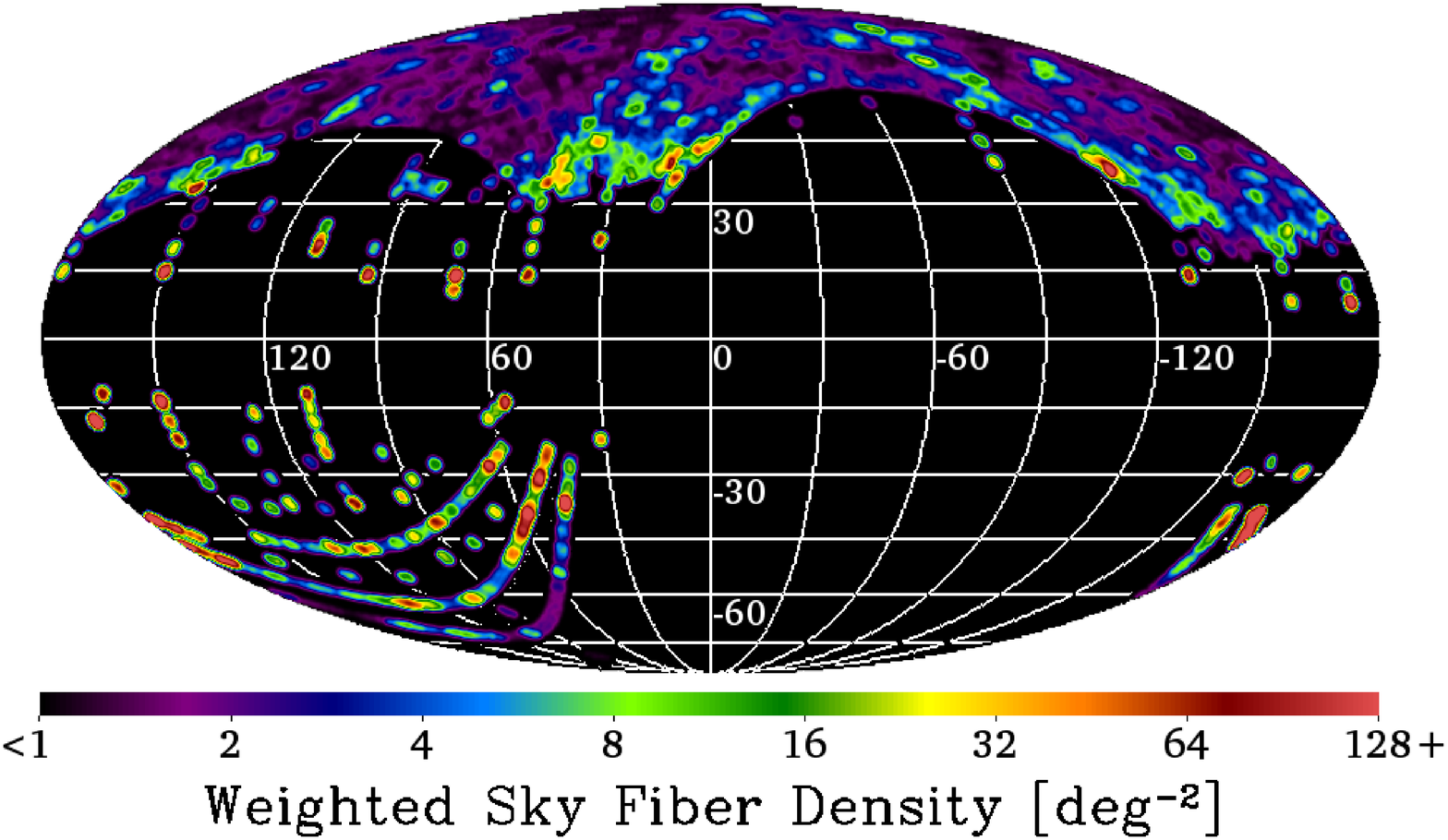}

\caption{Top panel: the sky coverage of the SDSS blank sky fibers in
  Galactic coordinates, centered around $(l, b) = (0,0)$.  Bottom
  panel: the sky coverage of the blank sky fibers weighted by
  $\sigma_{100\,\mu {\rm m,\,}p}^2 / \langle \sigma_{100\,\mu {\rm
  m}}^2 \rangle$ (Equation \eqref{eq:platevar}), which provides an
  approximate measure of each fiber's influence on the correlation
  spectrum.  The bulk of the SDSS footprint is near the North Galactic
  Pole, with additional concentrated sampling in Stripe 82 near the
  South Galactic Pole.  The useful survey footprint is much more
  heavily weighted toward equatorial regions.  }

\label{fig:skycoverage}

\end{figure}

Figure \ref{fig:skycoverage} shows the sky coverage of the SDSS
spectroscopic survey (upper panel), and the sky coverage weighted by
its power to affect the fit described by Equation
\eqref{eq:100mumodel} (lower panel).  For any plate, the effect on the
fit will be roughly proportional to the variance of the 100 $\mu$m
intensity over that plate,
\begin{equation}
\sigma_{100\,\mu {\rm m,\,}p}^2 \equiv \langle I_{100\,\mu {\rm m}}^2
\rangle_p - \langle I_{100\,\mu {\rm m}} \rangle_p^2~.
\label{eq:platevar}
\end{equation}
To obtain the weighted map, we have multiplied the fiber density by
the ratio of $\sigma_{100\,\mu {\rm m,\,}p}^2$ to its average over the
entire sample, $\langle \sigma_{100\,\mu {\rm m}}^2 \rangle$.  

Stripe 82 is clearly visible in the lower-left (and extreme
lower-right) parts of each map.  This region, defined by $-50^\circ <
{\rm RA} < 59^\circ$, $-1.25^\circ < \delta < 1.25^\circ$, was
repeatedly imaged as part of the SDSS Supernova Survey
\citep{Adelman-McCarthy+Agueros+Allam+etal_2006,
Frieman+Bassett+Becker+etal_2008,
Abazajian+Adelman-McCarthy+Agueros+etal_2009}.  It was also the target
of several ``special'' plates not part of the main SDSS surveys
\citep{Adelman-McCarthy+Agueros+Allam+etal_2006}.  Stripe 82, along
with a variety of patches at modest Galactic latitude and a swath at
the southern limit of SDSS's main footprint (in Galactic coordinates),
contributes the bulk of our measured signal.

\section{Results: Spectra} \label{sec:results_spec}

Here we present our correlation spectra of the DGL, $\alpha_\lambda$,
computed as described in Section \ref{sec:method}.  We present three
sets of spectra: the continuum of the scattered spectrum, and the
spectra from 4830-5040 \AA{} and from 6530-6770 \AA{} to show emission
lines.  We also divide the sky into three regions by Galactic latitude
and longitude to explore the spatial variation of the scattered light.

\subsection{Continuum of the Correlation Spectrum} \label{sec:dustcont}

We obtain the spectrum of the DGL associated with 100 $\mu$m emission
using the method described in Section \ref{sec:100u}.  By fitting
Equation (\ref{eq:100mumodel}) to the SDSS residual sky spectra, we
obtain a dimensionless coefficient $\alpha_\lambda$ at each wavelength
relating optical intensity to 100 $\mu$m emission.  We then mask
the nebular emission lines H$\alpha$, H$\beta$, [\textsc{N\,ii}]
$\lambda6550$, [\textsc{N\,ii}] $\lambda6585$, [\textsc{S\,ii}]
$\lambda6718$, and [\textsc{S\,ii}] $\lambda6733$ and bin the
correlation spectra in intervals of 50 \AA{} to increase the
signal-to-noise ratio.  We do not mask auroral lines like
[\textsc{O\,i}] $\lambda$6300, which are uncorrelated with the 100
$\mu$m intensity.  As we show in Section \ref{sec:errors}, our errors
are independent except for an interpolation effect, and they are
normally distributed.  The errors on the binned spectra pass the same
statistical test.

\begin{figure}
\includegraphics[width=\linewidth]{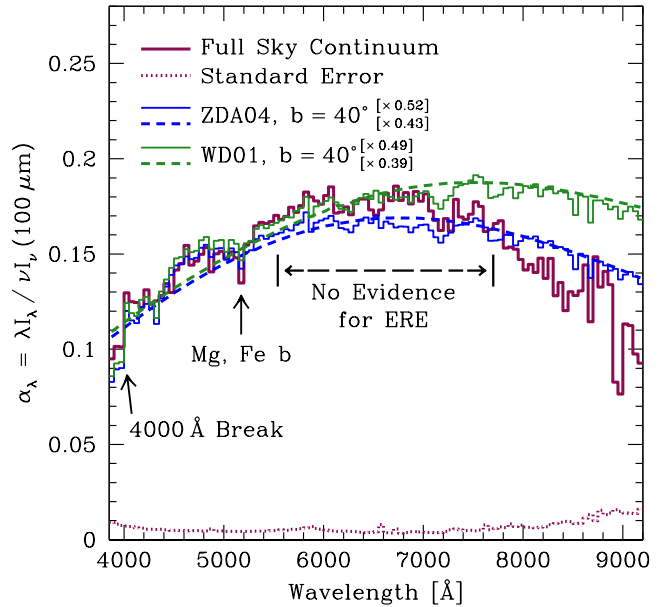}

\caption{The full-sky continuum of the scattered DGL, calculated as
  described in Section \ref{sec:method}.  We have masked nebular
  emission lines and binned the remaining spectrum into 50 \AA{}
  intervals.  The continuum is very blue and shows a clear 4000 \AA{}
  break and Mg and Fe b absorption just blueward of 5200 \AA; all of
  these features suggest scattered starlight.  Simplified radiative
  transfer calculations, described in Section \ref{sec:radtrans},
  confirm this and demonstrate that the DGL can discriminate between
  dust models.  Because our correlation spectrum is biased low by an
  unknown factor (Section \ref{sec:bias}), we have scaled all spectra
  to a common level.  The dashed lines use a featureless ISRF, while
  the solid lines use stellar population synthesis models.  The ZDA04
  model has fewer large grains and gives a better fit than WD01.  We
  find no evidence of Extended Red Emission (ERE) around 7000 \AA{}
  (Section \ref{sec:ere}).}
\label{fig:continuum_fullsky}
\end{figure}

Figure \ref{fig:continuum_fullsky} shows the continuum spectrum of the
DGL computed using all the sky fibers.  The spectrum is very blue, yet
it shows a clear 4000 \AA{} break characteristic of old stellar
populations.  Broad Mg and Fe b absorption are also visible just
blueward of 5200 \AA.  All of these characteristics are consistent
with a continuum of scattered starlight.  A simplified radiative
transfer calculation, discussed in Section \ref{sec:radtrans},
confirms scattering as the source of the DGL and shows that it may be
used to discriminate between dust models.  The four plotted curves use
two estimates of the continuum of the interstellar radiation field
(ISRF) and two dust models, the \cite{Zubko+Dwek+Arendt_2004} and
\cite{Weingartner+Draine_2001a} models (hereafter ZDA04 and WD01); an
excess of large grains in the latter produces a redder scattered
spectrum.  As we discuss in Section \ref{sec:bias}, errors and
small-scale structure in the 100 $\mu$m map bias our recovered
correlation spectrum low by an unknown factor, which we estimate to be
$2.1 \pm 0.4$ (see Section \ref{sec:bias}).  We have therefore scaled
the radiative transfer models to a common level to show their
different shapes.  The scale factors are indicated on the plot, and
are all consistent with our estimate for the bias.

\begin{figure*}
\includegraphics[width=0.5\linewidth]{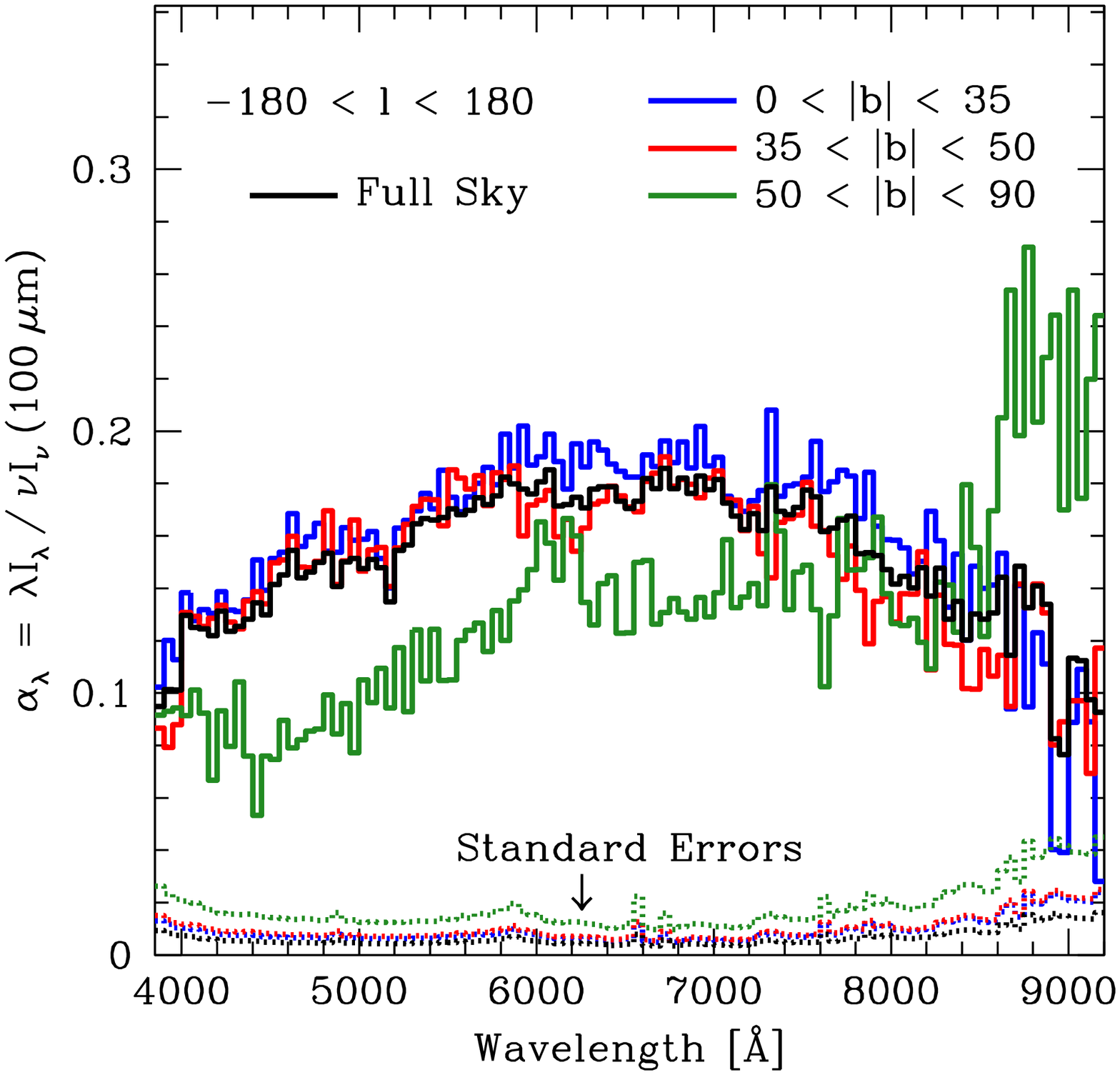}
\includegraphics[width=0.5\linewidth]{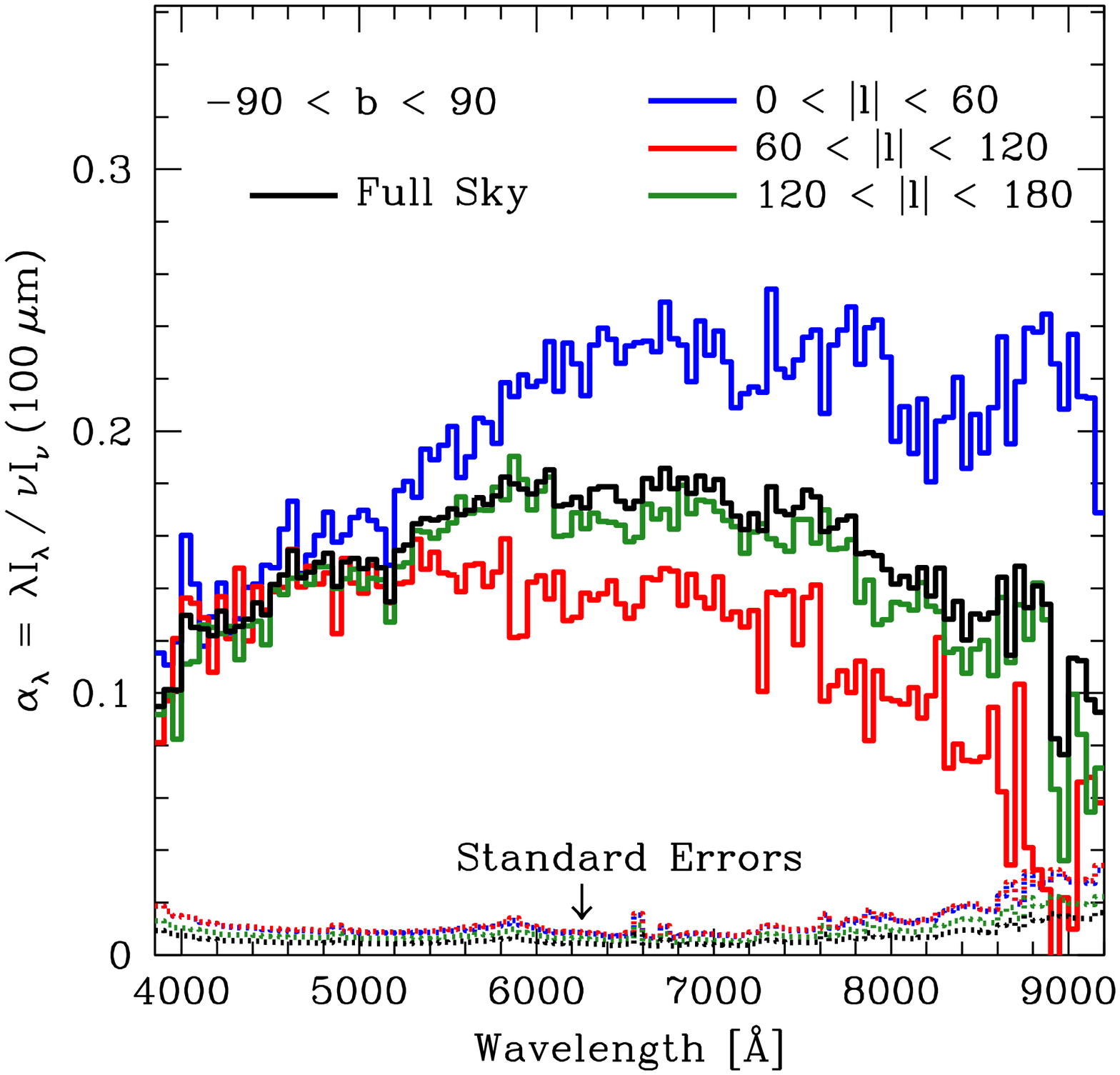}

\caption{The continuum spectra of the scattering component of the
  DGL for different regions of sky.  These are
  identical to Figure \ref{fig:continuum_fullsky} for restricted
  ranges of $l$ and $b$; note that we take $-180^\circ < l <
  180^\circ$.  The continuum shows some variation with Galactic
  latitude (left panel) and looks qualitatively different at the
  highest latitudes, potentially due to extragalactic contamination
  of the 100 $\mu$m map \citep{Yahata+Yonehara+Suto+etal_2007}.  The
  spectrum is also significantly redder toward the Galactic center
  (right panel).  This spatial variation could be due to differences
  in the dust composition, in the illuminating starlight, or
  uncorrected systematic effects.  }
\label{fig:continua_longlat}
\end{figure*}

Figure \ref{fig:continua_longlat} shows the continuum spectra from
restricted areas of the sky.  The left panel divides the sky by
Galactic latitude, with the ranges chosen to have comparable 100
$\mu$m emission summed over our sky fibers.  In the highest latitude
bin, this flux is distributed over a much larger number of sky fibers
and the resulting measurement is considerably noisier (see Figure
\ref{fig:skycoverage}).  Still, the errors are reliable (see Section
\ref{sec:errors}), and the shape differs from its values at lower
latitude by many sigma.  It is not clear whether the spatial variation
is due to variation in the dust properties, in the illuminating
starlight, systematic effects that become important when the signal is
weak, or some combination of these factors.  In particular,
\cite{Yahata+Yonehara+Suto+etal_2007} find a correlation of SDSS
galaxy density with SFD 100 $\mu$m emission in regions of low
extinction, consistent with extragalactic contamination of the SFD
map.  \cite{Yahata+Yonehara+Suto+etal_2007} report that this
contamination corresponds to an inferred $A_V \approx 0.01$, or
$I_{100\,\mu \rm m} \approx 0.2$ MJy\,sr$^{-1}$.  This would represent
about 15\% of the 100 $\mu$m intensity in our sky fibers with $|b| >
50^\circ$, and an even larger fraction of the intraplate variation to
which we are sensitive.

The right panel of Figure \ref{fig:continua_longlat} shows the
continuum spectra for different regions in Galactic longitude.  We
take the longitude $l$ to run from $-180^\circ$ to $180^\circ$, and
choose our regions to be equal in area and symmetric about the
Galactic center.  About half of our signal comes from the region
opposite the center, with $120^\circ < |l| < 180^\circ$.  As with
Galactic latitude, there are significant spatial variations in the
correlation spectrum.  It is unclear whether the redder spectrum in the
direction of the Galactic center is due to larger grains, redder
illuminating starlight, or some other effect.  This spectrum is the
most sensitive to the cutoff at high optical depth (Section
\ref{sec:100u}), which may indicate significantly reddened
illuminating starlight.

\subsection{Emission Lines} \label{sec:dustlines}

\begin{figure*}
\includegraphics[width=0.5\linewidth]{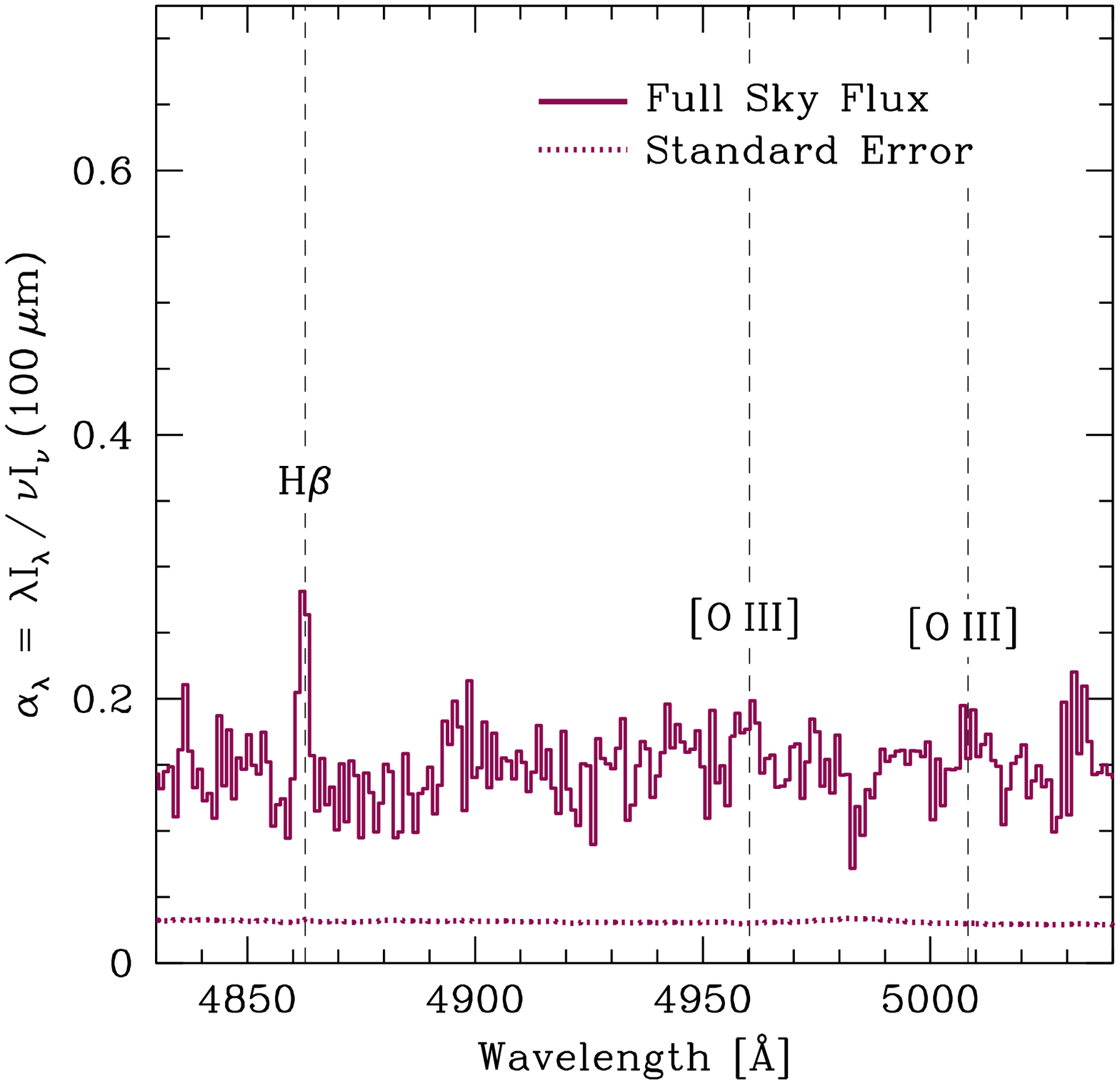}
\includegraphics[width=0.5\linewidth]{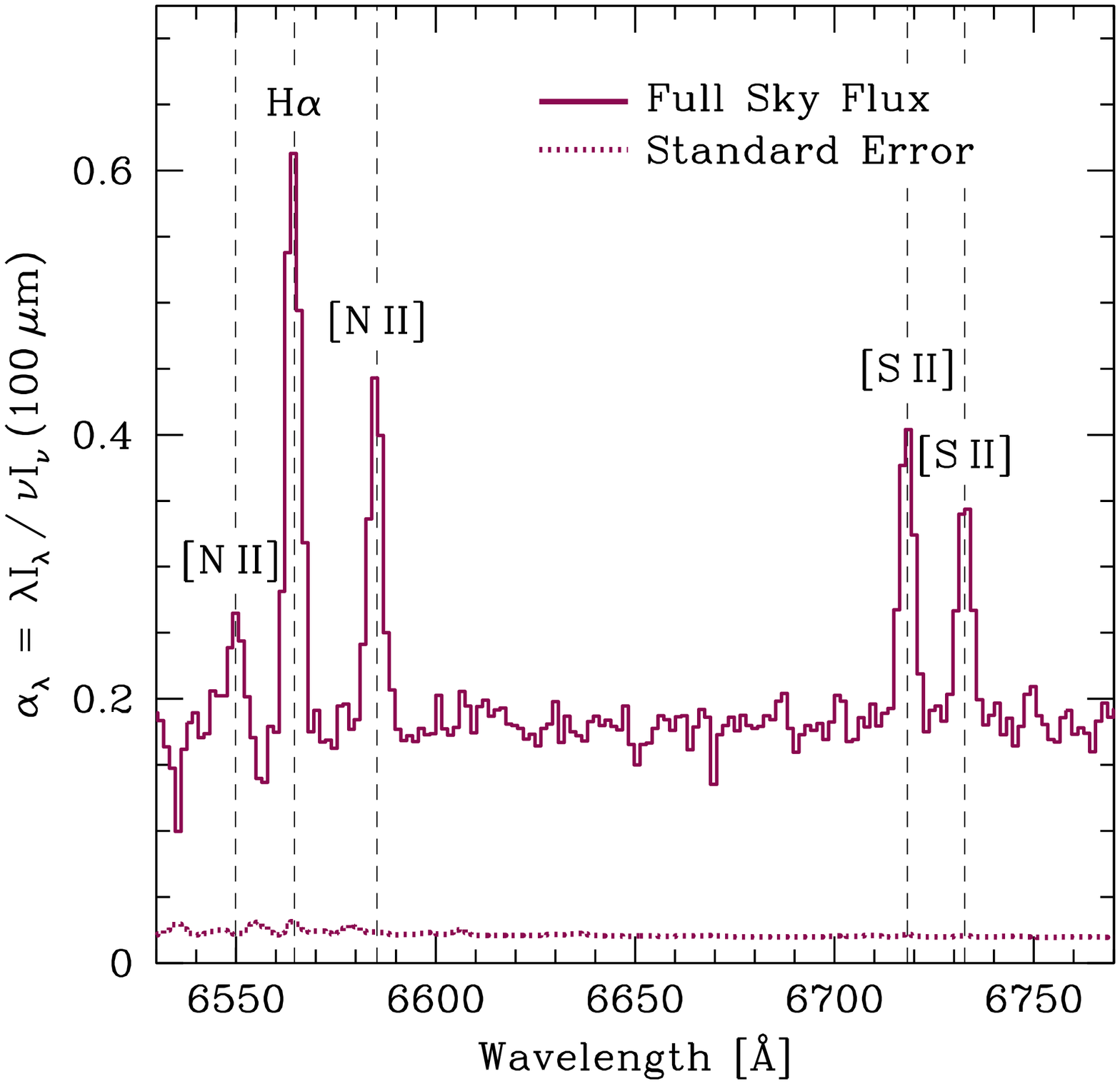}

\caption{Emission lines in the DGL.  H$\beta$,
  H$\alpha$, [\textsc{N\,ii}] $\lambda6550$, [\textsc{N\,ii}]
  $\lambda6585$, [\textsc{S\,ii}] $\lambda6718$, and [\textsc{S\,ii}]
  $\lambda6733$ are all detected with high significance, while we find
  little evidence of emission in [\textsc{O\,iii}] $\lambda5008$.  The
  lack of [\textsc{O\,iii}] indicates a relatively soft spectrum of
  ionizing photons compared to average \textsc{H\,ii} regions, while
  strong [\textsc{S\,ii}] and [\textsc{N\,ii}] indicate warm,
  $\sim$8,000 K gas.  }
\label{fig:emlines_fullsky}
\end{figure*}

Figure \ref{fig:emlines_fullsky} shows the correlation spectrum in the
wavelength ranges 4830-5040 \AA{} and 6530-6770 \AA{}, computed using
all of the sky fibers without binning the $\alpha_\lambda$.  The
spectrum of the DGL exhibits strong nebular emission lines which we
masked to show the continuum in Figures \ref{fig:continuum_fullsky}
and \ref{fig:continua_longlat}.  We detect H$\alpha$, H$\beta$,
[\textsc{N\,ii}] $\lambda6550$, [\textsc{N\,ii}] $\lambda6585$,
[\textsc{S\,ii}] $\lambda6718$, and [\textsc{S\,ii}] $\lambda6733$ at
high significance, but see only weak emission in the [\textsc{O\,iii}]
$\lambda5008$ line excited by early O-type stars.  Unfortunately, the
[\textsc{O\,ii}] $\lambda3727$ line lies just outside the SDSS
wavelength range.  These emission lines likely represent scattered
photons that were originally emitted from \textsc{H\,ii} regions and
the diffuse warm ionized medium (WIM).  

We list line strengths in Table \ref{tab:fullskylines} measured as
equivalent widths, which are unaffected by bias factors (Section
\ref{sec:bias}).  We use an interval of 6 SDSS wavelength elements,
corresponding to a velocity range of 400 km\,s$^{-1}$, to measure line
intensities.  For the lines between 6500 and 6800 \AA{}, we use the
average intensity over the interval from 6600 to 6700 \AA{} as our
estimate of the continuum.  For H$\beta$, we use 30 wavelength
elements on each side of the line, running from 4829-4858 \AA{} and
from 4864-4893 \AA.  The ratio of the [\textsc{N\,ii}] doublet,
[\textsc{N\,ii}] $\lambda6585$ to [\textsc{N\,ii}] $\lambda6550$,
provides a check on our recovered spectrum; our measured value of $2.8
\pm 0.6$ matches the ratio of 3 expected from the Einstein A
coefficients.

In order to measure the strengths of the Balmer lines in emission, we
need to account for the fact that they appear in absorption in stellar
spectra.  Fortunately, the strength of the Balmer absorption lines in
composite stellar spectra is closely correlated with that of the
$4000$ \AA{} Calcium break.  We measure the 4000 \AA{} break in our
continuum spectrum and use a range of model stellar spectra from
\cite{Bruzual+Charlot_2003}, hereafter BC03, to fit a linear
relationship between the strength of the break and the Balmer
equivalent widths.  We use models of 6 Gyr of constant star formation,
single stellar populations of 2.5 Gyr, 5 Gyr, and 11 Gyr, and two
exponential star formation histories, each with metallicities of 0.02
and 0.008 ($Z_\odot$ and $0.4Z_\odot$); a combination of these models
should provide a reasonable fit to stellar populations in the Solar
neighborhood.  Defining $\delta_{4000}$ to be the ratio of the
integrated intensity between 3850 and 4000 \AA{} to the integrated
intensity between 4000 and 4150 \AA{}, and computing the continua and
the line widths as described above, we find best-fit relationships of
\begin{align}
\left( \frac{{\rm EW}({\rm H}_\beta)}{\rm \AA} \right)
&\approx -2.2 \delta_{4000} + 0.17\quad {\rm
  and} \label{eq:eqwidth_hbeta} \\
\left( \frac{{\rm EW}({\rm H}_\alpha)}{\rm \AA} \right)
&\approx -1.5 \delta_{4000} - 0.19
\label{eq:eqwidth_halpha}
\end{align}
for the model stellar spectra.  The root-mean-square scatters of the
BC03 equivalent widths around the fits given by Equations
\eqref{eq:eqwidth_hbeta} and \eqref{eq:eqwidth_halpha} are $\sim$0.06
\AA{} and $\sim$0.1 \AA{}, respectively, significantly smaller than
the errors in our measurements of the DGL.  Correcting for stellar
absorption increases our measured H$\alpha$ and H$\beta$ line
strengths by $\sim$20\%.

\begin{deluxetable}{lccr}
\tablewidth{0pt}
\tablecaption{Equivalent Widths and Line Ratios in the DGL}
\tablehead{
  \colhead{Line} &
  \colhead{Equivalent Width [\AA]} &
  \multicolumn{2}{c}{Energy Ratio of Line to} \\
  \colhead{} &
  \colhead{} &
  \colhead{H$\alpha$} &
  \colhead{H$\beta$}
}
\startdata

H$\beta$ $\lambda4863$ & $4.8 \pm 0.7$\tablenotemark{a} & $0.38 \pm 0.05$ & 
1 \\ 
$[$\sc{O\,iii}$]$ $\lambda4960$ & $1.6 \pm 0.6$ & $0.14 \pm 0.05$ &
$0.38 \pm 0.15$ \\
$[$\sc{O\,iii}$]$ $\lambda5008$ & $0.8 \pm 0.6$ & $0.07 \pm 0.05$ &
$0.19 \pm 0.15$ \\
$[$He\sc{\,i}$]$ $\lambda5877$ & $0.3 \pm 0.8$ & $0.03 \pm 0.07$ &
$0.09 \pm 0.21$ \\
$[$\sc{N\,ii}$]$ $\lambda6550$ & $2.4 \pm 0.5$ & $0.19 \pm 0.04$ &
$0.50 \pm 0.12$ \\
H$\alpha$ $\lambda6565$ & $12.5 \pm 0.5$\tablenotemark{a} & 1 & 
$2.64 \pm 0.38$ \\
$[$\sc{N\,ii}$]$ $\lambda6585$ & $6.6 \pm 0.5$ & $0.53 \pm 0.04$ &
$1.40 \pm 0.22$ \\ 
$[$\sc{S\,ii}$]$ $\lambda6718$ & $5.7 \pm 0.4$ & $0.45 \pm 0.04$ &
$1.20 \pm 0.19$ \\
$[$\sc{S\,ii}$]$ $\lambda6733$ & $4.3 \pm 0.4$ & $0.34 \pm 0.04$ &
$0.91 \pm 0.15$

\enddata 
\tablenotetext{a}{Corrected for stellar absorption
using Equations \eqref{eq:eqwidth_hbeta} and \eqref{eq:eqwidth_halpha}.}
\label{tab:fullskylines}
\end{deluxetable}

The corrected equivalent width of H$\alpha$ in emission is consistent
with its value of 11 \AA{} in the local ISRF \citep[Table
12.1]{Draine_2011a}, obtained by integrating the full-sky H$\alpha$
map compiled by \cite{Finkbeiner_2003} from the Wisconsin H$\alpha$
Mapper \citep[WHAM,][]{Reynolds+Haffner+Madsen_2002}, Southern
H$\alpha$ Sky Survey Atlas
\citep[SHASSA,][]{Gaustad+McCullough+Rosing+VanBuren_2001}, and
Virginia Tech Spectral-Line Survey
\citep[VTSS,][]{Dennison+Simonetti+Topasna_1998}.  The line ratios
provide a probe of the average physical conditions in the local ISM.
The strength of the singly ionized [\textsc{N\,ii}] and
[\textsc{S\,ii}] lines and weakness of [\textsc{O\,iii}]$\lambda5008$
and [He\textsc{\,i}]$\lambda5877$ indicate that most of the S, N, and
O are singly ionized, while the He is largely neutral.  This implies a
lack of photons with $h\nu > 24.6$ eV in the Solar neighborhood, which
can be understood from the lack of stars of spectral type O8 and
earlier within 300 pc of the Sun.  The nearest eight O stars are
listed in Table \ref{tab:ostars}.  We discuss the physical conditions
of the local ISM in more detail in Section \ref{sec:discussion}.

\begin{deluxetable}{lccr}
\tablewidth{0pt}
\tablecaption{The Eight Nearest O Stars\tablenotemark{a}}
\tablehead{
  \colhead{GOS ID} &
  \colhead{Other ID} &
  \colhead{Spectral Type} &
  \colhead{Distance (pc)} 
}
\startdata

G006.28$+23.5901$ & $\zeta$ Oph     & O9.5V  & 112$^{+3}_{-3}$     \\
G203.86$-17.7401$ & $\delta$ Ori A  & O9.5V  & 221$^{+33}_{-25}$   \\
G206.45$-16.5901$ & $\zeta$ Ori A   & O9.7Ib & 239$^{+43}_{-32}$   \\
G202.94$+02.2001$ & 15 Mon          & O7V    & 309$^{+60}_{-43}$   \\
G255.98$-04.7101$ & $\zeta$ Pup     & O4I    & 335$^{+12}_{-11}$   \\
G262.80$-07.6901$ & $\gamma^2$ Vel  & O9:I:  & 349$^{+44}_{-35}$   \\
G195.05$-12.0001$ & $\lambda$ Ori A & O8III  & 361$^{+89}_{-60}$   \\
G206.82$-17.3401$ & $\sigma$ Ori AB & O9.5V  & 380$^{+136}_{-87}$ 

\enddata 
\tablenotetext{a}{References:
  \cite{Maiz-Apellaniz+Walborn+Galue+Wei_2004, Maiz-Apellaniz+Alfaro+Sota_2008}}
\label{tab:ostars}
\end{deluxetable}

\begin{figure*}
\includegraphics[width=0.5\linewidth]{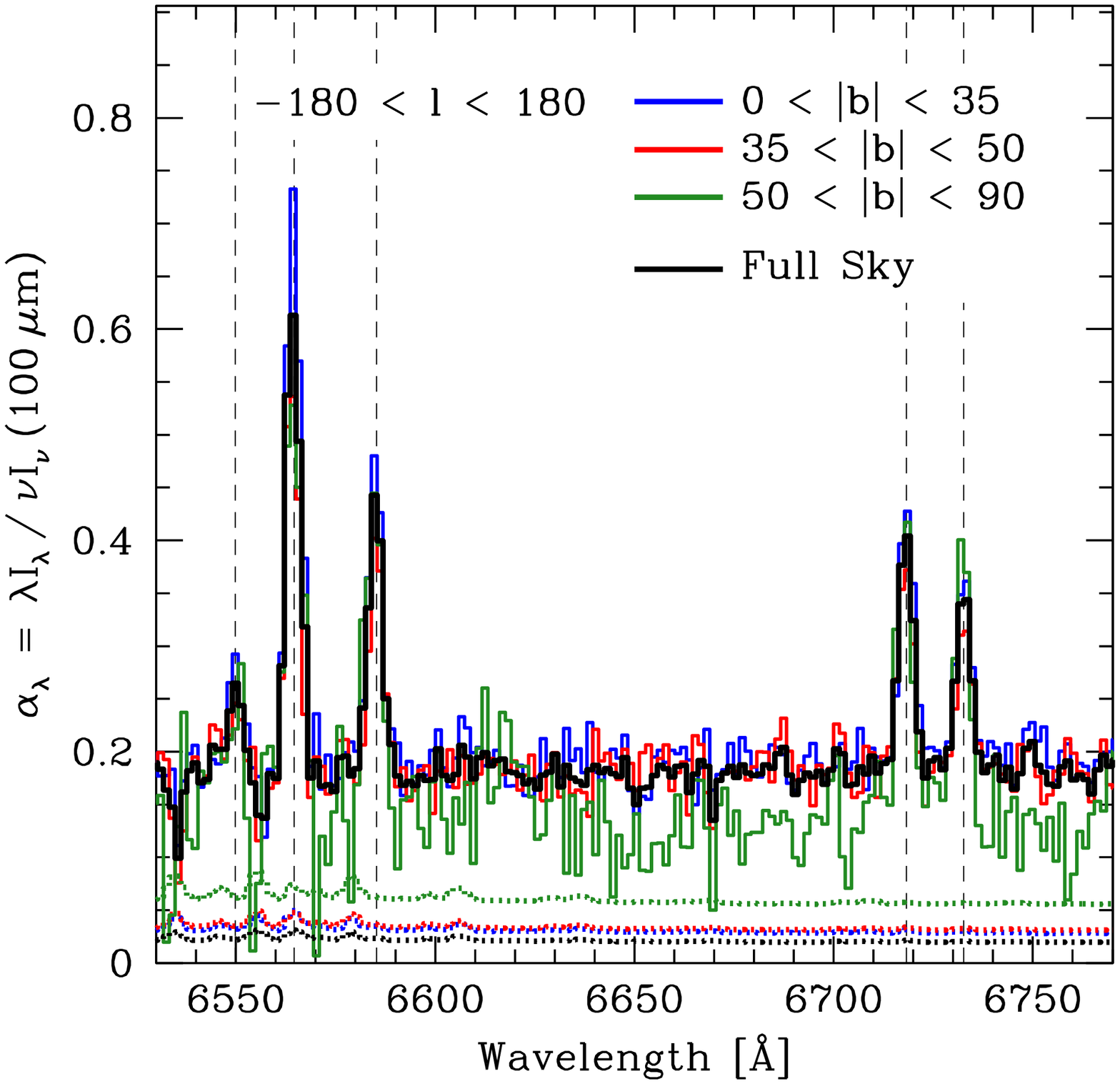}
\includegraphics[width=0.5\linewidth]{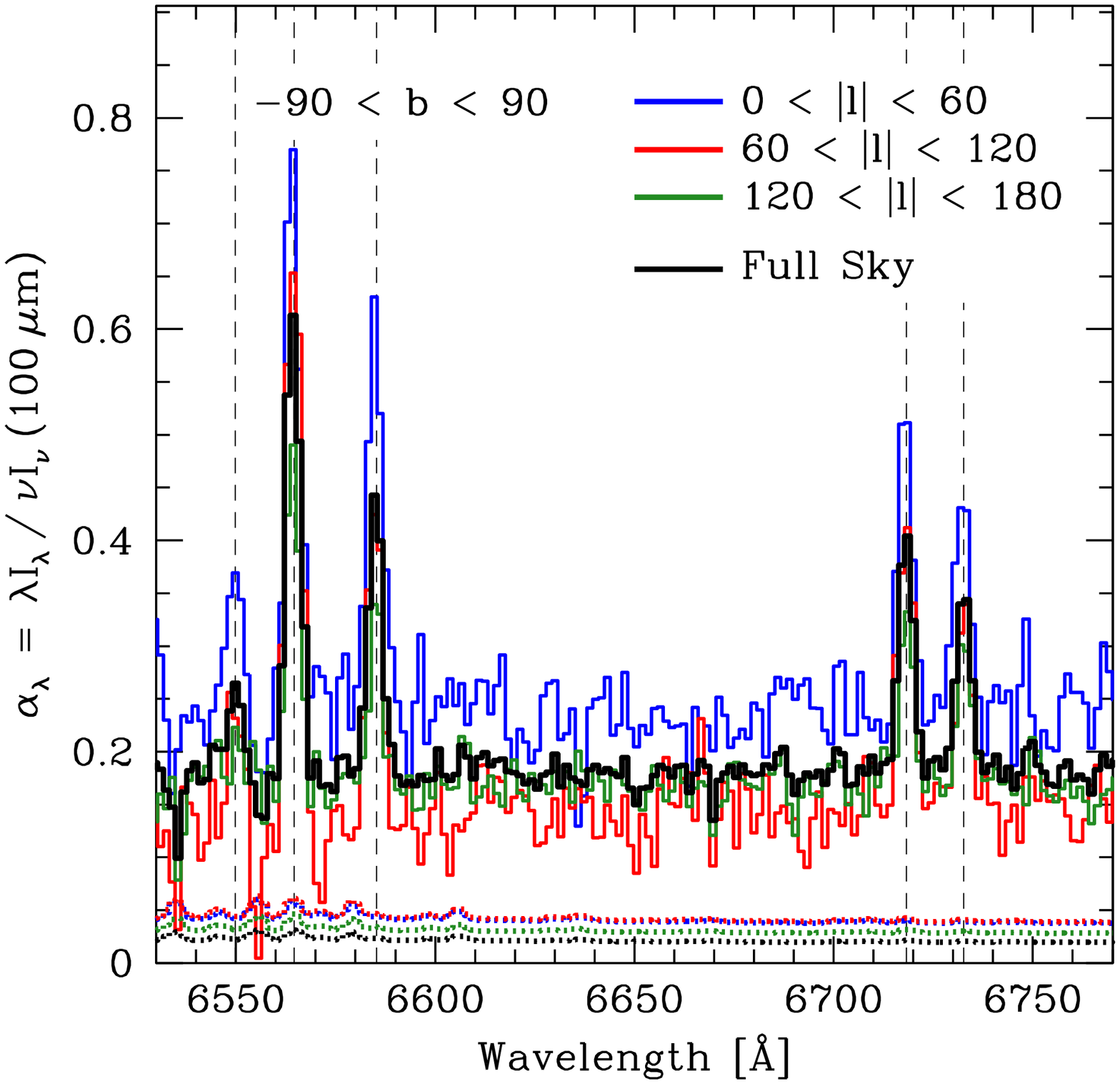}

\caption{Emission line spectra of the dust component of the DGL as
  functions of Galactic longitude and latitude.  The line ratios and
  equivalent widths depend only weakly on position, though the
  equivalent widths of all emission lines seem to decrease away from
  the Galactic center (green line, right panel).  }
\label{fig:emlines_longlat}
\end{figure*}

Figure \ref{fig:emlines_longlat} shows the strength of the emission
lines H$\alpha$, [\textsc{N\,ii}] $\lambda6550$, [\textsc{N\,ii}]
$\lambda6585$, [\textsc{S\,ii}] $\lambda6718$, and [\textsc{S\,ii}]
$\lambda6733$ relative to 100 $\mu$m emission for different ranges of
Galactic longitude and latitude.  The lines are somewhat stronger at
low Galactic latitude and in the direction of the Galactic center,
though this could reflect the relative number of nearby \textsc{H\,ii}
regions rather than the physical conditions in the ISM.  An important
caveat is that, because of spatial variations in the 100 $\mu$m
intensity, the correlation spectra in different regions of the sky
need not share the same bias factor (Sections \ref{sec:bias}
and \ref{sec:calib}).  The ratios of the [\textsc{S\,ii}] and
[\textsc{N\,ii}] lines to H$\alpha$, which are robust to calibration
difficulties, vary little across the sky.

\section{Calibration and Measurement Errors} \label{sec:errorbias}

The standard errors on our correlation spectra $\alpha_\lambda$ are
very nearly normally distributed with the variance given by our
maximum likelihood estimator, up to a constant factor due to our
interpolation of the original SDSS spectra onto a common wavelength
array.  The absolute calibration of the $\alpha_\lambda$ is more
problematic.  Our neglect of (unknown) measurement errors in the 100
$\mu$m intensity and structure unresolved by IRAS introduces a bias,
nearly independent of wavelength, which we estimate to be a factor of
$2.1 \pm 0.4$.  We demonstrate both of these results below.

\subsection{Measurement Errors} \label{sec:errors}

The errors in our correlation spectra are derived from fits to about
90,000 intensities over the full sky.  The sky spectra are first
interpolated onto a common wavelength array, introducing a correlation
between the intensities at neighboring wavelengths.  The difference
between the intensities $I_\lambda$ at wavelength elements $i$ and $i
+ 1$ ranges from $\frac{1}{2}$ to 1 time(s) its value without
interpolation (a factor of $\frac{1}{2}$ corresponds to the original
and final arrays being offset by half a wavelength increment).
Because the offsets between the original and resampled wavelengths are
random, the denominator of this fraction is uniformly distributed
between 1 and 2, which reduces the average difference $\langle I_{i +
1} - I_i \rangle$ by a factor of
\begin{equation}
\left< \frac{(I_{i + 1} - I_i)_{\rm orig}}{(I_{i + 1} - I_i)_{\rm
    interp}} \right> = \int_1^2 \frac{dt}{t} = \ln 2~.
\end{equation}
This correlation will propagate through the individual interpolated
spectra to the $\alpha_\lambda$.  If the element-to-element variations in
$\alpha_\lambda$ are dominated by Gaussian measurement errors, we
expect the quantities
\begin{equation}
\Delta \alpha_i \equiv \frac{\alpha_{i + 1} - \alpha_i}{(\ln 2)
  \sqrt{\sigma_{i+1}^2 + \sigma_i^2}}
\label{eq:normerrs}
\end{equation}
to be normally distributed with unit variance; $\alpha_i$ and
$\sigma^2_i$ are given by Equations \eqref{eq:maxlike} and
\eqref{eq:maxlikevar}, respectively.

\begin{figure}
\includegraphics[width=\linewidth]{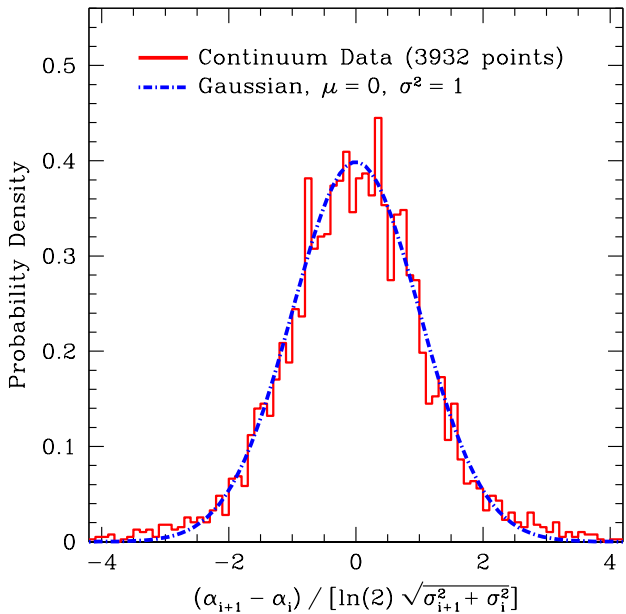}

\caption{Normalized difference between $\alpha_\lambda$ at successive
  wavelength elements (Equation \eqref{eq:normerrs}) in the full sky
  correlation spectrum, masking wavelengths near emission lines.  The
  factor of $\ln 2$ arises from the fact that we interpolate all
  spectra onto a common wavelength range, which introduces a
  correlation between neighboring wavelength elements (see Section
  \ref{sec:errors}).  There is no other scaling.  The resulting
  distribution of 3932 points is very nearly Gaussian; a K-S test
  rejects the consistency of the two distributions with just 83\%
  confidence.  }

\label{fig:specerrs}
\end{figure}

Figure \ref{fig:specerrs} shows that the distribution of $\Delta
\alpha_i$ as defined by Equation \eqref{eq:normerrs} is exceedingly
well-fit by a normal distribution with zero mean and unit variance.
We have masked 67 of the 3999 normalized intensity differences which
lie near the emission lines H$\alpha$, H$\beta$, [\textsc{N\,ii}]
$\lambda6550$, [\textsc{N\,ii}] $\lambda6585$, [\textsc{S\,ii}]
$\lambda6718$, and [\textsc{S\,ii}] $\lambda6733$.  At these
wavelengths, real spectral features contribute to the
element-to-element variation in $\alpha_\lambda$.  Even with 3932
values of $\Delta \alpha_i$ across the wavelength range, a
Kolmogorov-Smirnov (K-S) test only detects deviation from Gaussianity
with 83\% confidence.

The exceptional agreement shown in Figure \ref{fig:specerrs} gives us
confidence that our measured errors are reliable and independent (up
to a factor of $\ln 2$ from interpolating).  We have verified that
coadding neighboring wavelength elements, as we did to smooth the continuum in
Section \ref{sec:dustcont} and to compute the equivalent widths of
lines in Section \ref{sec:dustlines}, does not affect the statistical
properties of the errors.

\subsection{Biases in the Correlation Spectra} \label{sec:bias} 

Our model of the correlation between 100 $\mu$m intensity and optical
intensity (Equation \eqref{eq:100mumodel}) includes errors in the
SDSS sky fiber residuals, but neglects (unknown) errors in the 100
$\mu$m intensity at each fiber's location and structure unresolved by
IRAS.  This introduces a wavelength-independent bias to our recovered
spectra (Figures \ref{fig:continuum_fullsky} -
\ref{fig:emlines_longlat}).  We first derive an expression for the
bias and then estimate its value.

The correlation spectrum $\alpha_\lambda$ (Equation
\eqref{eq:100mumodel}) is derived from a $\chi^2$ minimization.  The
maximum likelihood values of the $\alpha_\lambda$ and their variances
are given by Equations \eqref{eq:maxlike} and \eqref{eq:maxlikevar},
with $x_{j, p}$ and $y_{\lambda, j, p}$ defined in Equations
\eqref{eq:ydef} and \eqref{eq:xdef}; $x_{j, p}$ is the excess 100
$\mu$m intensity in fiber $j$ relative to the average on plate $p$,
$y_{\lambda, j, p}$ is the residual sky fiber intensity $\lambda
I_\lambda$ at wavelength $\lambda$ in fiber $j$ on plate $p$, and
$\sigma^2_{\lambda, j, p}$ is its variance as estimated by the SDSS
pipeline.  We let $\xi_{j, p}$ denote the true excess 100 $\mu$m
emission at sky fiber $j$ on plate $p$, so that the measurement error,
including the effects of unresolved structure, is $\delta_{j, p} =
x_{j, p} - \xi_{j, p}$.

Assuming the model in Equation \eqref{eq:100mumodel} to be correct, we
may write $y_{\lambda, j, p}$ as $\alpha_\lambda \xi_{j, p} +
\varepsilon_{\lambda, j, p}$, with $\varepsilon_{\lambda, j, p}$
representing the measurement error in the sky fiber intensity.  We
further assume the error terms $\varepsilon_{\lambda, j, p}$ and
$\delta_{j, p}$ to be uncorrelated with zero mean and invoke the large
number of sky fibers (nearly $10^5$) to neglect sums of $\delta$ and
$\varepsilon$.  The first factor in Equation \eqref{eq:maxlike} becomes
\begin{equation}
\sum_{j, p} \frac{\left( y_{\lambda, j, p} \right) \left( x_{j, p}
  \right)}
{\sigma_{\lambda, y, j, p}^2} 
\approx \alpha_\lambda \sum_{j, p} \frac{\xi_{j, p}^2}
{\sigma_{\lambda, y, j, p}^2}~.
\end{equation}
This is the same value we would measure with $\delta_{j, p} = 0$.
Noting that the mean 100 $\mu$m excess, $\langle x_{j, p} \rangle$, is
zero by construction, our estimate of $\alpha_\lambda$ is biased low
by a factor
\begin{equation}
\frac{\alpha_\lambda(x, y)}{\alpha_\lambda(\xi, y)} \approx
\left( \sum_{j, p} \frac{\xi_{j, p}^2}{\sigma_{\lambda, y, j, p}^2} \right)
\left( \sum_{j, p} \frac{x_{j, p}^2}{\sigma_{\lambda, y, j, p}^2} \right)^{-1}
\approx \frac{\sigma^2_\xi}{\sigma^2_x}~,
\label{eq:bias}
\end{equation}
where $\sigma_x^2$ is the observed variance in $x$ and $\sigma_\xi^2
\approx \sigma_x^2 - \langle \delta^2 \rangle$ would be its value with
no measurement errors.  Thus,
\begin{equation}
\frac{\alpha_\lambda(x, y)}{\alpha_\lambda(\xi, y)} \approx
1 - \frac{\langle \delta^2 \rangle}{\sigma_x^2}~.
\label{eq:bias2}
\end{equation}
Note that this bias is independent of wavelength and of the errors in
the sky fiber intensities.  We have empirically verified the latter by
adding noise to the intensities; we recover the same correlation
spectra to within the errors.

\subsection{Calibrating the Correlation Spectra} \label{sec:calib} 

An unbiased estimator for $\alpha_\lambda$, for example a likelihood
function of the form
\begin{equation}
\mathcal{L}(x_j, y_{\lambda, j} | \alpha_\lambda)
= \int \mathcal{L}(x_j | x) \, 
\mathcal{L}(y_{\lambda, j} | x, \alpha_\lambda)\,
\mathcal{L}(x) \, dx~,
\end{equation}
would remove all calibration issues.  The SDSS pipeline provides an
estimate of $\mathcal{L}(y_{\lambda, j} | x, \alpha_\lambda)$, the
likelihood of measuring sky fiber residual $y_{\lambda}$ in sky fiber
$j$ given a correlation spectrum $\alpha_\lambda$ and true 100 $\mu$m
intensity $x$.  Unfortunately, we have almost no information on the
errors in the 100 $\mu$m intensity to estimate $\mathcal{L}(x_j | x)$,
the likelihood of measuring excess 100 $\mu$m intensity $x_j$ given a
true value $x$, and can only guess at a prior, $\mathcal{L}(x)$, from
the 100 $\mu$m map itself.

Because of these difficulties, we use two alternative and independent
approaches.  We first construct an estimator that we expect to be
asymptotically unbiased, assuming the error in residual 100 $\mu$m
intensity to depend weakly on the true residual value at a fiber
position.  This is a reasonable assumption, particularly because
we are subtracting the mean 100 $\mu$m emission over a plate; an
$x$-value of zero does {\it not} correspond to zero intensity.  We
then show the results of our radiative transfer calculations assuming
ZDA04 dust and a plane-parallel galaxy.  

\begin{figure}

\includegraphics[width=\linewidth]{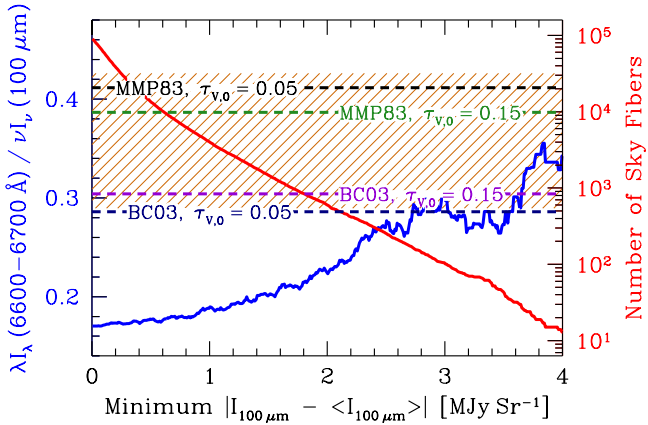}

\caption{The correlation spectrum $\alpha_\lambda$ between 6600 and
  6700 \AA{} using only points with large excess 100 $\mu$m
  intensities.  The red curve indicates the number of sky fibers
  satisfying each 100 $\mu$m cutoff.  If the errors on residual 100
  $\mu$m intensity depend weakly on fiber position, this estimator
  will be asymptotically unbiased.  We overplot several
  single-scattering radiative transfer models, which we describe in
  detail in Section \ref{sec:radtrans}.  All assume ZDA04 dust, but
  with different dust densities and different ISRFs.  We
  conservatively adopt the bias factor of $2.1 \pm 0.4$ indicated by
  the orange shading to calibrate our correlation spectrum.  }

\label{fig:bias}
\end{figure}

Under the assumption of uniform errors in the residual 100 $\mu$m
intensity, Equation \eqref{eq:bias2} shows that restricting the sample
to fibers with large $|x|$ (and therefore large $\sigma^2_x$) will
give an asymptotically unbiased estimate.  We therefore recalculate
$\alpha_\lambda$ between 6600 and 6700 \AA{} using only the fibers
with $|x| > |x|_{\rm min}$.  Figure \ref{fig:bias} shows
$\alpha_\lambda$ as a function of $|x|_{\rm min}$, calculated by
varying $|x|_{\rm min}$ from 0 to 4 MJy\,sr$^{-1}$; as expected,
$\alpha_\lambda$ increases with $|x|_{\rm min}$.  Figure
\ref{fig:bias} suggests a bias factor of at least $\sim$1.7$-$2.  This
is also supported by our radiative transfer model, discussed in detail
in Section \ref{sec:radtrans}, which suggests a bias of
$\approx$1.7$-$2.4.  We conservatively adopt a bias factor of $2.1 \pm
0.4$, indicated by the shaded region of Figure \ref{fig:bias}, to
calibrate our correlation spectrum.

\section{Discussion} \label{sec:discussion}

\subsection{Scattered Light and Dust Models} \label{sec:radtrans}

Many features of the DGL -- the 4000 \AA{} break, broad Mg $+$ Fe b
absorption, and a much bluer continuum than that of stars with these
spectral features -- support the hypothesis that the DGL is dominated
by scattered starlight.  This is hardly surprising, as the spectrum
was derived by correlating residual optical intensity with 100 $\mu$m
intensity over small spatial scales.  A simplified radiative transfer
calculation confirms that our spectrum is consistent with dust
scattering and allows us to use the DGL to discriminate between dust
models.

Our model of the DGL uses an infinite plane-parallel galaxy with a
Gaussian vertical distribution of dust,
\begin{equation}
\rho_{\rm dust} \propto \exp \left[ -\frac{z^2}{2 \sigma^2} \right]~,
\end{equation}
with $\sigma = 250$ pc \citep{Malhotra_1995, Nakanishi+Sofue_2003}.
We approximate the stellar distribution as the sum of two exponential
distributions with scale heights of 300 pc and 1350 pc
\citep{Binney+Merrifield_1998, Gilmore+Reid_1983}.  The 300 pc
component dominates the distribution with about 90\% of the stars.  

We use two estimates of the stellar emission spectrum:
\begin{enumerate}
\item A model that reproduces the local ISRF of
  \cite{Mathis+Mezger+Panagia_1983}, hereafter MMP83, in the midplane,
  and
\item A stellar population synthesis model from BC03, with solar
  metallicity and an exponential star formation history over 12 Gyr.
\end{enumerate}
The MMP83 ISRF for $\lambda > 2460$ \AA{} is approximated as a sum of
three dilute blackbodies of $T = 3000$, 4000, and 7000 K, with
dilution coefficients $W = 5\times 10^{-13}$, $1.65 \times 10^{-13}$,
and $10^{-14}$, respectively.  We compute the attenuation of the
stellar emission using Equation \eqref{eq:powerpt}, setting the {\it
attenuated} spectrum equal to the local ISRF.  In this way, we
``de-redden'' the MMP83 ISRF to obtain the stellar source spectrum.  

\begin{figure}
\includegraphics[width=\linewidth]{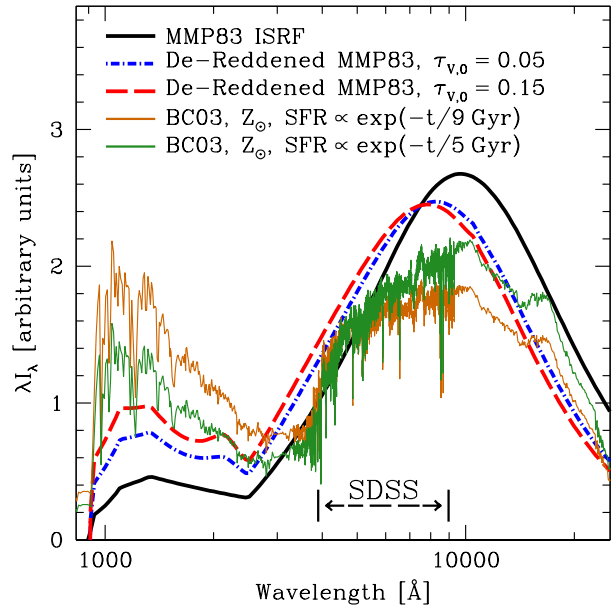}
\caption{Comparison of the MMP83 ISRF, de-reddened using our radiative
  transfer models, and BC03 models with solar metallicity and
  exponential star formation histories over 12 Gyr.  The BC03 models
  have significant excesses in the UV that increase dust heating and
  reduce the ratio of the optical scattering to far-infrared
  emission.}
\label{fig:isrf_compare}
\end{figure}

\begin{figure}
\includegraphics[width=\linewidth]{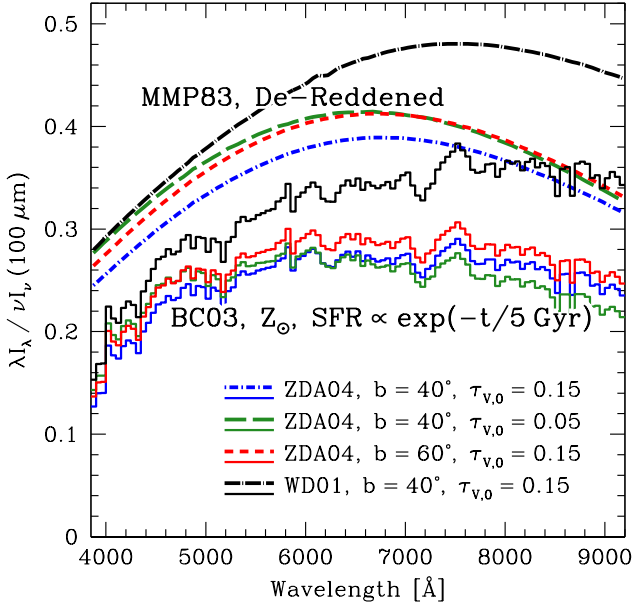}
\caption{Single-scattering radiative transfer calculations assuming a
  plane-parallel galaxy.  The results are much more sensitive to the
  assumed dust model than to the parameters of the galaxy model.  The
  WD01 model produces more scattering at longer wavelengths than the
  ZDA04 model because of an excess of large dust grains.  }
\label{fig:modelspec}
\end{figure}

We show all of the ISRF spectra in Figure \ref{fig:isrf_compare}.  The
peak of the de-reddened MMP83 $\lambda F_\lambda$ is at about 8000
\AA{}, corresponding to a temperature of $\sim$4600 K.  The spectrum
is similar to a BC03 model with a 5 Gyr star formation timescale,
though MMP83 has less UV emission.  This is probably a result of the
relative lack of early O stars in the Solar neighborhood and of our
model's neglect of extinction from dust in a young star's birth cloud
\citep{Charlot+Fall_2000}.  The UV discrepancy becomes much more
serious for BC03 models with more extended star formation histories.
This UV excess significantly increases the dust heating and decreases
the ratio of scattering in the optical to emission in the
far-infrared.  For a star formation timescale of 5 Gyr, this is a
$\sim$30-40\% effect relative to the de-reddened MMP83 model (see
Figures \ref{fig:bias} and \ref{fig:modelspec}).

Theoretical and observational estimates of the local star formation
history favor roughly constant star formation rates over $\sim$10 Gyr
\citep{Hernandez+Avila-Reese+Firmani_2001,
Cignoni+DeglInnocenti+Prada_Moroni+etal_2006}.  Such models do not
agree with our measured spectrum of the DGL and, using our radiative
transfer model, would produce a very different ISRF from MMP83.  This
may indicate that a substantial fraction of the illuminating starlight
originates relatively far from the Solar neighborhood, it may be a
result of our simplified radiative transfer, or it may indicate that
stars in the Solar neighborhood are generally older than is currently
thought.  The spatial variation of the DGL (Figure
\ref{fig:continua_longlat}) shows that the geometry of the radiative
transfer problem is likely to be important.  A more detailed model of
the Galaxy could better constrain the average stellar source spectrum.

Our radiative transfer calculations neglect multiple scatterings.
This is a good approximation at high Galactic latitude where the
optical depths are low.  It becomes poor near the midplane, but SDSS
has very little sky coverage near the Galactic plane (Figure
\ref{fig:skycoverage}).  We assume all absorbed starlight to be
reradiated isotropically in the infrared and use a Henyey-Greenstein
phase function for scattering in the optical.  We use the dust model
of \cite{Draine+Li_2007} to convert total infrared power to IRAS 100
$\mu$m bandpass power, with
\begin{equation}
\left( \nu I_\nu \right)_{100\,\mu \rm m}
= (0.52 \pm 0.05) I_{\rm TIR}~.
\label{eq:nuInu/TIR}
\end{equation}
The central value corresponds to their model with an incident
starlight intensity 80\% of that in the Solar neighborhood, which
\cite{Draine+Li_2007} found to provide the best fit to the average
far-infrared spectrum measured by
\cite{Finkbeiner+Davis+Schlegel_1999}.  The confidence interval in
Equation \eqref{eq:nuInu/TIR} includes models from 0.5 to 1.5 times
the local starlight intensity.  Once a dust model is specified, with
wavelength-dependent cross-sections, albedos, and anisotropy
parameters, our model for the scattered light spectrum has no free
parameters.  We derive the relevant equations in the Appendix and
evaluate them numerically.

Figure \ref{fig:modelspec} shows the results of our calculations to be
remarkably insensitive to the details of the galaxy modeling (other
than the assumed stellar source spectrum).  Though not shown, models
with a single exponential distribution of stars and with an
exponential, rather than Gaussian, dust distribution, are nearly
indistinguishable from the present models.
For $\csc |b| > 1.4$ ($|b|< 45^\circ$) the \textsc{H\,i} has $N_{\rm
H}\approx 2.9\times10^{20}\csc |b|~{\rm cm}^{-2}$
\citep{Dickey+Salpeter+Terzian_1978} falling below this relation for
$b> 45^\circ$.  For $E(B-V)/N_{\rm H} = 5.8\times10^{21}~{\rm
cm}^2\,{\rm mag}^{-1}$ and $R_V=A_V/E(B-V)=3.1$, we would then expect
$\tau_V=0.17\csc |b|$ for $|b|<45^\circ$, and lower values at
$b>45^\circ$.  We use $\tau_V = 0.15 \csc |b|$ for our fiducial
distribution.

\begin{figure}
\includegraphics[width=\linewidth]{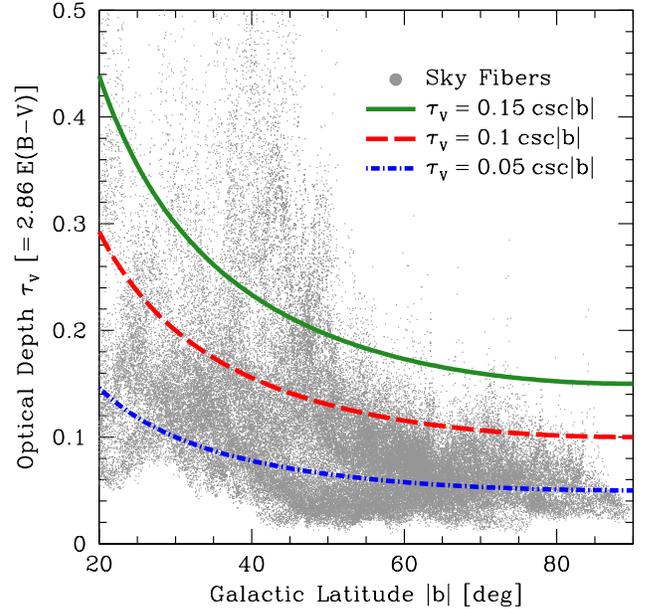}
\caption{Optical depths at each sky fiber computed using the SFD
estimates of $E(B-V)$ and assuming $R_V=3.1$ dust.  The dust
distribution is roughly bracketed by $\tau_V = 0.05 \csc |b|$ and
$\tau_V = 0.15 \csc |b|$; these two models predict nearly the same
spectrum of scattered light.  }
\label{fig:dustdist}
\end{figure}

Figure \ref{fig:dustdist} shows that the optical depths at the SDSS
sky fibers, computed using the SFD estimates of $E(B-V)$ and assuming
$R_V=3.1$ dust, are roughly bracketed by $\tau_V = 0.05 \csc |b|$ and
$\tau_V = 0.15 \csc |b|$.  This suggests that the sky fiber locations
were slightly biased toward lower-than-average \textsc{H\,i} column
densities.

While the details of the model galaxy have little effect on the
predicted spectrum, the dust model matters a great deal.  The WD01
model appears to have too many large grains, giving too much
scattering at long wavelengths.  The size distribution of ZDA04 brings
our model into much better agreement with the data (see Figure
\ref{fig:continuum_fullsky}).  While our model predicts the scattering
spectrum to be a very weak function of Galactic latitude, variations
in the dust properties, geometry, or illuminating starlight could
result in much larger differences.  Possible spatial variations in 100
$\mu$m errors and small-scale structure (Section \ref{sec:bias})
further complicate the interpretation of the spatial variations seen
in Figure \ref{fig:continua_longlat}.

\subsection{Line Emission} \label{sec:lines}

On the evidence presented above, we can be confident that the
continuum of the correlation spectrum $\alpha_\lambda$ consists
primarily of scattered starlight.  It is more difficult to show that
the line emission is scattered rather than from ionized gas physically
associated with the dust.  The strongest piece of evidence is that the
equivalent width of H$\alpha$ in the correlation spectrum, $12.5 \pm
0.5$ \AA{}, matches its measured value of 11 \AA{} in the local ISRF
(Section \ref{sec:dustlines}).  Because scattering will preserve the
equivalent width of H$\alpha$ in the ISRF incident on the dust, a
stronger H$\alpha$ line would have indicated an additional, nearly
continuum-free component seen in direct emission and correlating with
100 $\mu$m intensity.

If the emission lines observed in Section \ref{sec:dustlines} are
observed mostly or entirely in reflection, they provide a probe of the
average physics of the nearby ISM.  As discussed in
Section \ref{sec:dustlines}, the strength of the [\textsc{N\,ii}] and
[\textsc{S\,ii}] lines, combined with the weakness of the
[\textsc{O\,iii}] lines,
indicate relatively few photons with $h \nu > 24.6$ eV.
This is probably due to the lack of early-type O stars in the Solar
neighborhood.
\begin{figure}
\includegraphics[width=\linewidth]{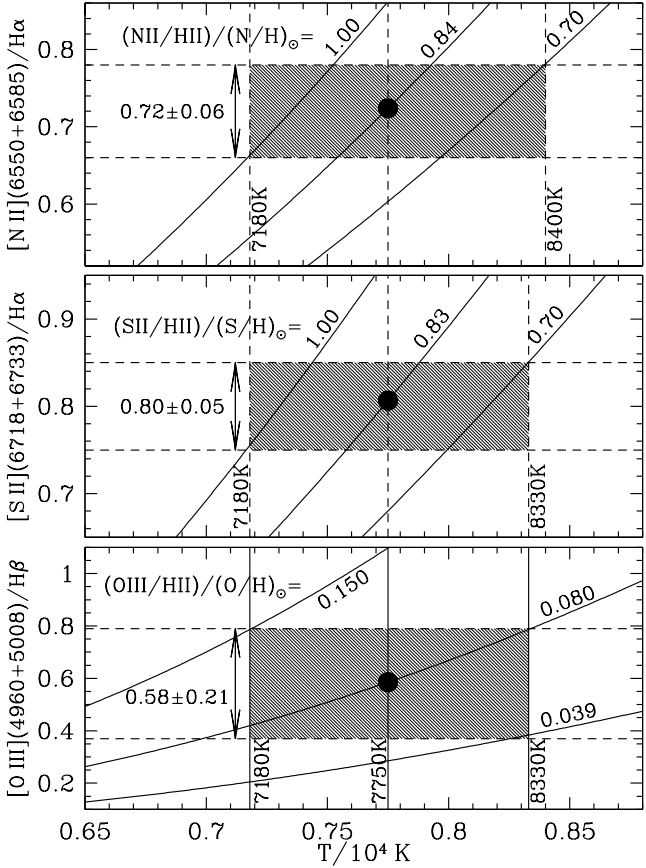}
\caption{
Upper panel: [\textsc{N\,ii}](6550+6585)/H$\alpha$ vs.\ $T$.
Curves are predicted values for indicated values of 
\textsc{N\,ii}/\textsc{H\,ii} relative to (N/H)$_\odot$.
If $0.7 < $(\textsc{N\,ii}/\textsc{H\,ii})/(O/N)$_\odot < 1$, then the
observed line ratio $0.72\pm0.06$ requires $7180<T<8400$K.
Middle panel: [\textsc{S\,ii}](6718+6733)/H$\alpha$ vs.\ electron
temperature $T$.  Curves are predicted values for indicated
values of \textsc{S\,ii}/\textsc{H\,ii} relative to (S/H)$_\odot$.
The observed line ratio $0.80\pm0.05$ 
and $0.7<$(\textsc{S\,ii}/\textsc{H\,ii})/(O/H)$_\odot < 1$
limits the temperature to
$7180 < T < 8330$K.
Lower panel: \textsc{O\,iii}(4960+5008)/H$\beta$ vs. $T$.
Curves are predicted values for indicated values of
\textsc{O\,iii}/\textsc{H\,ii} relative to (O/H)$_\odot$.
The observed line ratio $0.58\pm0.21$ and the allowed 
temperature range $7180 < T < 8330$K implies that \textsc{O\,iii}/\textsc{H\,ii}
is between 0.039$\times$ and 0.15$\times$(O/H)$_\odot$.
The dot in each panel indicates our best estimate:
$T=7700$K, 
\textsc{N\,ii}/\textsc{H\,ii}=0.85(N/H)$_\odot$,
\textsc{S\,ii}/\textsc{H\,ii}=0.83(S/H)$_\odot$, and
\textsc{O\,iii}/\textsc{H\,ii}=0.08(O/H)$_\odot$.
}
\label{fig:line ratios}
\end{figure}
The observed strengths of the collisionally-excited lines 
relative to recombination lines of similar wavelength are: 
\begin{align}
\frac{[\rm S\,\textsc{ii}](6718 + 6733)}{\rm H \alpha~ 6565} &= 0.80 \pm 0.05~,
\\
\frac{[\rm N\,\textsc{ii}](6550 + 6585)}{\rm H \alpha~ 6565} &= 0.72 \pm
0.06~,~{\rm and} \\
\frac{[\rm O\,\textsc{iii}](4960 + 5008)}{\rm H \beta~ 4863} &= 0.58 \pm 0.21
~.
\end{align}
These allow us to estimate the temperature of the ISM where the lines originate 
and the state of ionization of S, N, and O.
Figure \ref{fig:line ratios} shows predicted line ratios as functions
of electron temperature $T$, calculated using H$\alpha$ and H$\beta$
emissivities from \citet{Draine_2011a},
collision strengths 
for \textsc{N\,ii} from \citet{Hudson+Bell_2005},
for \textsc{S\,ii} from \citet{Tayal+Zatsarinny_2010}, and
for \textsc{O\,iii} from \citet{Aggarwal+Keenan_1999}.
A density $n_e=10^2\,{\rm cm}^{-3}$ was assumed; the results are insensitive
to $n_e$ provided $n_e\lesssim 10^3\,{\rm cm}^{-3}$.

Nitrogen is not depleted in the ISM, and the first and second ionization
potentials (14.0 and 29.6 eV) lead us to expect \textsc{N\,ii}/\textsc{H\,ii}
to be close to the solar abundance (N/H)$_\odot=7.4\times10^{-5}$
\citep{Asplund+Grevesse+Sauval+Scott_2009}.
If we assume $0.7<$(\textsc{N\,ii}/\textsc{H\,ii})/(N/H)$_\odot<1$
we see from Figure \ref{fig:line ratios} that 
the observed
[\textsc{N\,ii}](6550+6565)/H$\alpha$ ratio
allows only temperatures $7180<T<8400$ K.

Sulfur is not expected to be depleted in \textsc{H\,ii} regions or the
diffuse ISM.  With an ionization potential of 23.38 eV for
\textsc{S\,ii}$\rightarrow$\textsc{S\,iii}, \textsc{S\,ii} will be the
dominant ionization stage in \textsc{H\,ii} regions where He is
neutral.  If we assume \textsc{S\,ii}/\textsc{H\,ii} to be between 0.7
and 1.0$\times$ the solar abundance (S/H)$_\odot=1.45\times10^{-5}$
\citep{Asplund+Grevesse+Sauval+Scott_2009}, then we see from Figure
\ref{fig:line ratios} that the observed
[\textsc{S\,ii}(6718+6733)]/H$\alpha$ requires $7180 < T < 8330$ K.

Oxygen is only slightly depleted in the diffuse ISM, with $\sim$20\%
of the O resident in silicates.  The second ionization potential of
oxygen is 35.1~eV, and therefore \textsc{O\,iii} will be present only
when He is ionized.  A star of spectral type O8 or earlier is required
for the He ionization zone to account for more than 50\% of the mass
in the \textsc{H\,ii} region \citep{Draine_2011a}.  As seen in Table
\ref{tab:ostars}, the nearest such stars are 15 Mon (O7V, $d =
309^{+60}_{-43}$ pc) and $\zeta$ Pup (O4I, $d = 335^{+12}_{-11}$ pc).
The reflected light in the DGL is expected to originate mainly within
a few hundred pc of the Sun, and therefore the contribution from
\textsc{H\,ii} regions should be dominated by \textsc{H\,ii} regions
where He is neutral and O is singly ionized.  Figure \ref{fig:line
ratios} shows that the observed strength of
[\textsc{O\,iii}](4960+5008)/H$\beta$ is consistent with
\textsc{O\,iii}/\textsc{H\,ii} between 0.039 and 0.15 of (O/H)$_\odot
= 5.4\times10^{-4}$ \citep{Asplund+Grevesse+Sauval+Scott_2009}.  The
observed [\textsc{O\,iii}] emission can be reproduced by emission from
\textsc{H\,ii} regions with $T\approx 7700$ K and
\textsc{O\,iii}/\textsc{H\,ii} $\approx 0.08\times ({\rm O}/{\rm
H})_\odot$.

%
He\textsc{\,i}~5877 is not detected, with a 3-$\sigma$ upper limit
He\textsc{\,i}~5877/H$\alpha < 0.24$; for $T\approx7700$ K, this
corresponds to He\textsc{\,ii}/\textsc{H\,ii} $<0.48$, which is not a
useful constraint.

The emission lines measured in the correlation spectrum
$\alpha_\lambda$ are therefore consistent with the line ratios
expected for \textsc{H\,ii} regions within $\sim$400~pc of the Sun,
for electron temperature $T\approx 7700\pm250$ K.  These temperatures
and ionization states are also consistent with the WIM
\citep{Madsen+Reynolds+Haffner_2006}, which we expect to contribute a
significant fraction of the line emission, particularly at high
latitudes.

\subsection{The Fraction of Scattered H$\alpha$ at High Latitude}

The previous section argued that the H$\alpha$ in our correlation
spectrum is scattered.  We can then use the calibrated correlation
spectra to estimate the fraction of
the observed H$\alpha$ at high Galactic latitude that is scattered light.
We simply integrate the correlation spectrum over
the H$\alpha$ line and multiply by the ratio of the total 100 $\mu$m to
H$\alpha$ emission over the high latitude sky fibers.  

From Figures \ref{fig:bias} and \ref{fig:modelspec}, we take the ratio
$\lambda I_\lambda (6600\,{\rm \AA}) / \nu I_\nu (100\,\mu$m$)$ to be
$0.38 \pm 0.07$.  This continuum value allows us to convert an
H$\alpha$ equivalent width in our correlation spectrum into an
intensity relative to the 100 $\mu$m intensity.  We then have
\begin{equation}
\label{eq:Halpha/100um}
\frac{I({\rm H} \alpha)}{\rm R} \approx 
(0.090 \pm 0.017) \times
\left( \frac{W_\lambda({\rm H}\alpha )}{12.5\,{\rm \AA}} \right) \times
\left( \frac{I_\nu (100\,\mu{\rm m})}{\rm MJy\,sr^{-1}} \right),
\end{equation}
where the H$\alpha$ intensity $I({\rm H}\alpha)$ is in
Rayleighs\footnote{$1\,{\rm R}=\frac{10^6}{4\pi}$
photons\,cm$^{-2}$\,s$^{-1}$\,sr$^{-1}$} and we recall (Table
\ref{tab:fullskylines}) that $W_\lambda({\rm H}\alpha)\approx
12.5{\,\rm \AA}$ in our correlation spectrum.  Our result
(\ref{eq:Halpha/100um}) is close to the value $I({\rm H}\alpha)/{\rm
R} = (0.129\pm0.015)(I_\nu(100\,\micron)/{\rm MJy\,sr^{-1}})$ found
recently by \citet{Witt+Gold+Barnes+etal_2010}.

Averaging the 100 $\mu$m intensity from SFD and the
H$\alpha$ intensity from \cite{Finkbeiner_2003} over all regions with
$b > 60^\circ$ gives a ratio 
\begin{equation}
\label{eq:total halpha/total 100}
{\bigg\langle} \frac{I({\rm H}\alpha)}{\rm R} {\bigg\rangle}
 = 0.47
{\bigg\langle}\frac{I_\nu(100\,\micron)}{\rm MJy\,sr^{-1}}{\bigg\rangle}
~.
\end{equation}

The ratio of the coefficients in (\ref{eq:Halpha/100um},\ref{eq:total
halpha/total 100}) gives a scattered fraction of H$\alpha$ at high
latitudes of $0.090/0.47 = 0.19$.  We add the statistical error of 5\%
in the H$\alpha$ equivalent width and our conservative estimate of the
uncertainty in the calibration shown in Figure \ref{fig:bias} in
quadrature, giving a scattered H$\alpha$ fraction at high latitude of
\begin{equation}
\label{eq:scattered fraction}
\frac{{\rm scattered ~H}\alpha}{{\rm total~H}\alpha} = 0.19 \pm 0.04
~.
\end{equation}
\citet{Witt+Gold+Barnes+etal_2010} produced a scatter diagram of the
scattered H$\alpha$ fraction (their Fig.~6).  The centroid of their
distribution appears to be close to our measured value $0.19\pm0.04$ .

\citet{Wood+Reynolds_1999} estimated that 5--20\% of the H$\alpha$ at
high latitudes would be scattered light from \textsc{H\,ii} regions.
In a theoretical study of the emission spectrum of the diffuse H$\alpha$,
\citet{Dong+Draine_2011} concluded that the observed line ratios,
including the low ratio of radio free-free to H$\alpha$, could 
be understood if $\sim$20\% of the diffuse H$\alpha$ is
actually reflected from dust, rather than emission from recombining gas
in that direction.
All of these results are consistent with our value $0.19\pm0.04$.

We expect the scattered H$\alpha$ to originate both from
\textsc{H\,ii} regions around young stars and from the WIM.  Our line
ratios seem to be more consistent with those of the WIM than with
classical \textsc{H\,ii} regions \citep{Madsen+Reynolds+Haffner_2006},
which may indicate that the WIM dominates the line emission in our
correlation spectra.  However, this could also be due to the lack of
nearby \textsc{H\,ii} regions illuminated by early O stars.  Our line
ratios do seem to be compatible with those in the $\phi$ Per
\textsc{H\,ii} region, which is illuminated by a B0.5$+$sdO system
\citep[Fig.~12 of][]{Madsen+Reynolds+Haffner_2006}.

\subsection{Dust Luminescence} \label{sec:ere}

We find no evidence of the Extended Red Emission (ERE) observed in
some reflection nebulae (e.g., NGC 7023) as a broad emission excess
peaking near 7000\AA, with a FWHM $\sim$1500\AA\ \citep[][and
references therein]{Witt+Vijh_2004}; a similar broad excess is
reported to be present in the diffuse ISM
\citep{Gordon+Witt+Friedmann_1998,Szomoru+Guhathakurta_1998,
Witt+Mandel+Sell+etal_2008}.  The spectrum of light reflected by
cirrus clouds was measured by \citet{Szomoru+Guhathakurta_1998}, who
reported a broad peak in $\lambda I_\lambda$ near 6500\AA\ that was
attributed to luminescence, on the grounds that scattering is
insufficient.  Our study also finds $\lambda I_\lambda$ peaking near
6500\AA\ (see Figure \ref{fig:continuum_fullsky}) but the observed
spectrum appears to be consistent with what is expected for
scattering.  Indeed, our scattering models tend to {\it over}-predict
the DGL around 7000$-$9000 \AA{} (Figure
\ref{fig:continuum_fullsky}), depending on the adopted dust size
distribution.

If we posit that the slight excess emission between 5500 \AA{} and
7500 \AA{} over our prediction with the ZDA04
dust model (see Fig.\ \ref{fig:continuum_fullsky}) is an upper limit
on the ERE, we infer that dust luminescence accounts for no more than
10\% of the DGL in this wavelength range and less elsewhere.  After
calibrating the spectra and relating $I_\nu(100\,\micron)$ to the total
IR power using Eq.\ (\ref{eq:nuInu/TIR}), this implies a ratio of ERE
to infrared power of $\lesssim$0.4\%, or a ratio of ERE to scattered
power of $\lesssim$1\%.  

The ERE will be a function of the intensity of illuminating UV light
and the dust column density.  It is possible that our technique, which
requires spatial variations of the ERE correlated with 100 $\mu$m
intensity over a $\sim$1$^\circ$ scale, is ill-suited to its
detection.  If the UV ISRF is particularly weak or spatially variable
on small scales, it could at least partially explain the lack of ERE
in our correlation spectrum.  We have therefore performed two tests.

First, we made our optical depth cutoff $A_V \lesssim 0.5$ more
stringent to restrict the calculation to regions optically thin to UV
photons.  Reducing this cutoff by a factor of 2 had a negligible
effect on the correlation spectrum.  Unfortunately, when we reduced it
further, the main stellar spectral features disappeared and new
features similar to those in the high latitude spectrum of Figure
\ref{fig:continua_longlat} appeared.  At such low levels of 100 $\mu$m
emission, we expect extragalactic contamination to be significant.  

We have also tried to estimate empirically how much of the small-scale
variation (over $\sim$1$^\circ$) in 100 $\mu$m emission is due to
variations in the illuminating starlight and how much is due to
variations in the dust column density.  We recalculated the
correlation spectrum, correlating the SDSS intensities against dust
optical depth rather than 100 $\mu$m intensity.  Our recovered
correlation spectrum was nearly identical.  Unfortunately, the
resolution of the SFD temperature map used to convert from 100 $\mu$m
intensity to optical depth, $0.7^\circ$, is too coarse to make this
test conclusive.  

While we find no evidence of hidden ERE in our correlation spectrum, a
more thorough investigation may require instruments that can more
directly probe the dust-scattered optical spectrum.

\section{Summary and Conclusions}

In this paper, we have measured the spectrum of the scattered
component of the DGL using 90,000 calibration spectra from SDSS by
correlating against 100 $\mu$m intensity measured by COBE and IRAS and
reduced by SFD.  The correlation spectrum of the DGL is consistent
with scattering of photons emitted by stars and by ionized gas.  Its
continuum and lines show interesting variations across the sky, which
could be due to differences in the illuminating starlight, in the dust
properties, or structure in the Galaxy.  Our spectrum is not
calibrated because the correlation neglects (unknown) errors and
small-scale structure in the 100 $\mu$m intensity.  However, an
asymptotically unbiased estimator agrees well with simplified
radiative transfer calculations and allows us to calibrate the average
spectrum of the DGL.  In calibrating the data, we also provide an
estimate of the errors and small-scale structure in the SFD 100 $\mu$m
map.

With a simplified radiative transfer calculation assuming a
plane-parallel galaxy, we show that the spectrum of the DGL can
discriminate between dust models.  The ZDA04 dust model is preferred
to the WD01 model, which has more large grains and scatters too much
red light.  The ZDA04 model fits the data very well without requiring
dust luminescence as a source of ERE.  Indeed, our radiative transfer
calculations tend to over-predict scattering redward of 7000 \AA.

Our measurement of H$\alpha$ in the DGL allows us
to indirectly measure the fraction of scattered H$\alpha$ at high
Galactic latitude.  We find an average value of $0.19 \pm 0.04$,
consistent with the results of \citet{Wood+Reynolds_1999},
\citet{Witt+Gold+Barnes+etal_2010} and the value inferred by
\cite{Dong+Draine_2011}.  The line emission also allows us to
constrain the properties of local \textsc{H\,ii} regions.  The lack of
emission from species with high ionization energies is consistent with
the lack of nearby early O stars, and the average scattered spectrum
is consistent with gas at $\sim$7500 K.

Our results are the product of the extraordinary SDSS, the only
dataset of its kind.  It would be possible, though difficult, to
incorporate the new Baryon Oscillations Spectroscopic Survey spectra
\citep[BOSS,][]{Eisenstein+etal_2011} into the analysis.  The new
fibers are smaller (2$''$ instead of 3$''$), losing far more light and
making spectrophotometric calibration much more difficult.  The DGL is
a sufficiently weak signal that it requires thousands of blank sky
spectra to extract useful results.  These spectra also need to be in
regions with significant dust columns, regions that tend to be avoided
by large extragalactic surveys.  The SDSS spectra will likely remain
the best tool for analyzing the DGL for some time.

While the optical data are unlikely to improve in the near future, WISE
and AKARI may allow significant improvements to the infrared map.  Our
results indicate that the measurement errors and small-scale structure
in the SFD 100 $\mu$m map are nearly as large as the $\sim$1$^\circ$
variations to which our correlations are sensitive.  By combining high
resolution infrared data with the SFD maps, we could improve the
quality of our DGL spectra and resulting
constraints on Galactic stars and dust.  

\acknowledgments{We wish to thank Michael Strauss, Dan Maoz, Yasushi
  Suto, and Steve Bickerton for many helpful conversations, and
  Michael Strauss for numerous suggestions that greatly improved the
  manuscript.  We also wish to thank Jim Gunn and Robert Lupton for
  their help understanding the intricacies of SDSS data.  This
  material is based upon work supported by the National Science
  Foundation Graduate Research Fellowship under Grant No.~DGE-0646086.
  This research was supported in part by NSF grant AST-1008570.

  Funding for the SDSS and SDSS-II has been provided by the Alfred
  P. Sloan Foundation, the Participating Institutions, the National
  Science Foundation, the U.S. Department of Energy, the National
  Aeronautics and Space Administration, the Japanese Monbukagakusho,
  the Max Planck Society, and the Higher Education Funding Council for
  England. The SDSS Web Site is \verb|http://www.sdss.org/| .


  }

\appendix

\section{Scattering and Absorption in a Plane-Parallel Exponential
  Galaxy}

We test dust models by calculating scattered spectra in an infinite
plane-parallel galaxy, using vertical distributions of stars and dust
appropriate to the Solar neighborhood.  The Sun is assumed to lie in
the midplane, $z = 0$.  We calculate the absorption (and hence,
infrared emission) and scattering along each line-of-sight, neglecting
multiple scatterings.

We denote the (wavelength-dependent) optical depth of dust above
height $z$ along a vertical line-of-sight as $\tau_\lambda(z)$, 
\begin{equation}
\tau_\lambda(z) \equiv \int_z^{\infty} \sigma_{\rm ext}(\lambda)
 \rho(z')\,dz'~,
\label{eq:optdepth}
\end{equation}
where $\sigma_{\rm ext}$ is the extinction cross-section and $\rho$ is
the dust density.  Consider a sheet of stars with uniform surface
power density $\Sigma_\lambda$ at height $z_{\rm s}$.  The optical
depth from an annulus of stars at a distance $R$ from the grain's
projected position in the stellar sheet to a grain at height $z$ is
\begin{equation}
A_\lambda(z, z_{\rm s}, R) 
\equiv |\tau_\lambda(z) - \tau_\lambda(z_{\rm s})|
\frac{\sqrt{(z - z_{\rm s})^2 + R^2}}{|z - z_{\rm s}|}~.
\label{eq:optext}
\end{equation}
The total flux density (neglecting multiply scattered photons)
incident on a grain at height $z$ is then
\begin{align}
F_{\lambda}(z, z_{\rm s}) &= \int_0^\infty 
2 \pi R\,dR \frac{\Sigma_\lambda\,
\exp \big[ -A_\lambda(z, z_{\rm s}, R) \big]}
{4\pi \left[(z - z_{\rm s})^2 + R^2
    \right]} \nonumber \\
&= \left( \frac{\Sigma_\lambda}{2} \right) 
E_1 \left(|\tau_\lambda(z) - \tau_\lambda(z_{\rm s}) | \right)~,
\label{eq:powerpt}
\end{align}
where $E_1$ is the exponential integral and $\tau_\lambda$ is related
to $z$ by Equation \eqref{eq:optdepth}.  The total reradiated
intensity is an integral of the absorbed fraction of Equation
\eqref{eq:powerpt} over all wavelengths and stellar sheets.  Defining
$\omega_\lambda$ as the albedo, the total infrared intensity $I_{\rm
TIR}$ from a sightline at Galactic latitude $b$ is
\begin{equation}
I_{\rm TIR} (b) = \frac{\csc |b|}{8 \pi}
\int_0^\infty \left( 1 - \omega_\lambda \right) d\lambda 
\int_{-\infty}^{\infty} \frac{d \Sigma_\lambda}{dz_{\rm s}}\, dz_{\rm s}
\int_0^{\tau_\lambda (0)} d\tau' \, E_1 \left( |\tau' -
\tau_\lambda(z_{\rm s}) | \right)~.
\label{eq:infrared}
\end{equation}
A dust model, like the \cite{Draine+Li_2007} model used in Equation
\eqref{eq:nuInu/TIR}, is needed to convert $I_{\rm TIR}$ into a specific
intensity $I_{\nu}$ at infrared frequency $\nu$.

Anisotropic scattering makes it more difficult to calculate the
optical intensity.  Consider an annulus of the stellar sheet centered
directly below a dust grain, and let $\theta$ sweep out the annulus,
with $\theta = 0$ pointed away from us.  If the dust grain is along a
sightline at Galactic latitude $b$, the law of cosines gives the
required scattering angle $\xi$ as
\begin{equation}
\cos \xi = 
 \frac{(z - z_{\rm s})^2 - R z \cos \theta \cot b}
     {\sqrt{\left(z^2 \cot^2 b + (z - z_{\rm s})^2 \right)
 \left( R^2 + (z - z_{\rm s})^2 \right) }}~.
\label{eq:scatang}
\end{equation}
The intensity of light scattered in our direction from a stellar sheet
at height $z_{\rm s}$ is then given by
\begin{equation}
I_{\lambda,\,\rm sca}(z, z_{\rm s}, b) 
= \int_0^\infty R\,dR \int_0^{2 \pi} d\theta\,
\phi_\lambda \left( \cos \xi \right) 
\frac{\Sigma_\lambda\,\exp \big[ -A_\lambda(z, z_{\rm s}, R) \big]}
{4 \pi \left((z - z_{\rm s})^2 + R^2 \right)}~,
\end{equation}
with $\phi_\lambda$ being the normalized phase function and
$A_\lambda$ defined by Equation \eqref{eq:optext}.  We use a
Henyey-Greenstein phase function for simplicity.  If $\phi$ is
constant, the $\theta$ integral is trivial and we recover Equation
\eqref{eq:powerpt} modulo a factor of $4\pi$.  Scattered light is
further attenuated by dust along the line-of-sight.  Neglecting
multiple scatterings, we finally have
\begin{equation}
I_{\lambda,\,\rm sca}(z_{\rm s}, b) 
= \omega_\lambda \csc|b| \int_0^{\tau_\lambda (0)}
\exp \left[-\csc|b| \left( \tau_\lambda (0) - \tau' \right) \right]\,d\tau'
\int_0^\infty R\,dR
\int_0^{2 \pi} d\theta\,\phi_\lambda \left( \cos \xi \right)
\frac{\Sigma_\lambda\,\exp \big[ -A_\lambda(z, z_{\rm s}, R) \big]}
{4 \pi \left((z - z_{\rm s})^2 + R^2 \right)}~,
\label{eq:scatpower}
\end{equation}
where $z$ and $\tau_\lambda$ are related by Equation
\eqref{eq:optdepth} and $A_\lambda$ is defined by Equation
\eqref{eq:optext}.  As for Equation \eqref{eq:infrared}, we integrate
Equation \eqref{eq:scatpower} over $z_{\rm s}$.  We do not integrate
over wavelength; the spectrum of $I_{\lambda,\, \rm sca}$ is shown in
Figure \ref{fig:continuum_fullsky} normalized by the infrared
intensity $\nu I_\nu$ at 100 $\mu$m, which is related to $I_{\rm TIR}$
by Equation \eqref{eq:nuInu/TIR}.  The ratio
$I_{\lambda,\, \rm sca}/I_{\rm TIR}$ depends very weakly on Galactic
latitude $b$, the normalization of the optical depth, and the details
of the stellar and dust distributions.  Once an ISRF and dust model
are supplied, there are no other free parameters.

\bibliographystyle{apj_eprint}
\bibliography{btdrefs}

\begin{thebibliography}{}

\bibitem[\protect\citeauthoryear{{Abazajian} et~al.}{{Abazajian}
  et~al.}{2009}]{Abazajian+Adelman-McCarthy+Agueros+etal_2009}
{Abazajian}, K.~N., {Adelman-McCarthy}, J.~K., {Ag{\"u}eros}, M.~A., et~al.
  2009, \apjs, 182, 543

\bibitem[\protect\citeauthoryear{{Adelman-McCarthy} et~al.}{{Adelman-McCarthy}
  et~al.}{2008}]{Adelman-McCarthy+Agueros+Allam+etal_2008}
{Adelman-McCarthy}, J.~K., {Ag{\"u}eros}, M.~A., {Allam}, S.~S., et~al. 2008,
  \apjs, 175, 297

\bibitem[\protect\citeauthoryear{{Adelman-McCarthy} et~al.}{{Adelman-McCarthy}
  et~al.}{2006}]{Adelman-McCarthy+Agueros+Allam+etal_2006}
{Adelman-McCarthy}, J.~K., {Ag{\"u}eros}, M.~A., {Allam}, S.~S., et~al. 2006,
  \apjs, 162, 38

\bibitem[\protect\citeauthoryear{{Aggarwal} \& {Keenan}}{{Aggarwal} \&
  {Keenan}}{1999}]{Aggarwal+Keenan_1999}
{Aggarwal}, K.~M.,  \& {Keenan}, F.~P. 1999, \apjs, 123, 311

\bibitem[\protect\citeauthoryear{{Asplund} et~al.}{{Asplund}
  et~al.}{2009}]{Asplund+Grevesse+Sauval+Scott_2009}
{Asplund}, M., {Grevesse}, N., {Sauval}, A.~J.,  \& {Scott}, P. 2009, \araa,
  47, 481

\bibitem[\protect\citeauthoryear{{Binney} \& {Merrifield}}{{Binney} \&
  {Merrifield}}{1998}]{Binney+Merrifield_1998}
{Binney}, J.,  \& {Merrifield}, M. 1998, {Galactic Astronomy} (Princeton, NJ:
  Princeton University Press)

\bibitem[\protect\citeauthoryear{{Boggess} et~al.}{{Boggess}
  et~al.}{1992}]{Boggess+Mather+Weiss+etal_1992}
{Boggess}, N.~W., {Mather}, J.~C., {Weiss}, R., et~al. 1992, \apj, 397, 420

\bibitem[\protect\citeauthoryear{{Bruzual} \& {Charlot}}{{Bruzual} \&
  {Charlot}}{2003}]{Bruzual+Charlot_2003}
{Bruzual}, G.,  \& {Charlot}, S. 2003, \mnras, 344, 1000

\bibitem[\protect\citeauthoryear{{Charlot} \& {Fall}}{{Charlot} \&
  {Fall}}{2000}]{Charlot+Fall_2000}
{Charlot}, S.,  \& {Fall}, S.~M. 2000, \apj, 539, 718

\bibitem[\protect\citeauthoryear{{Cignoni} et~al.}{{Cignoni}
  et~al.}{2006}]{Cignoni+DeglInnocenti+Prada_Moroni+etal_2006}
{Cignoni}, M., {Degl'Innocenti}, S., {Prada Moroni}, P.~G.,  \& {Shore}, S.~N.
  2006, \aap, 459, 783

\bibitem[\protect\citeauthoryear{{Dennison}, {Simonetti}, \&
  {Topasna}}{{Dennison} et~al.}{1998}]{Dennison+Simonetti+Topasna_1998}
{Dennison}, B., {Simonetti}, J.~H.,  \& {Topasna}, G.~A. 1998, \pasa, 15, 147

\bibitem[\protect\citeauthoryear{{Dickey}, {Terzian}, \& {Salpeter}}{{Dickey}
  et~al.}{1978}]{Dickey+Salpeter+Terzian_1978}
{Dickey}, J.~M., {Terzian}, Y.,  \& {Salpeter}, E.~E. 1978, \apjs, 36, 77

\bibitem[\protect\citeauthoryear{{Dong} \& {Draine}}{{Dong} \&
  {Draine}}{2011}]{Dong+Draine_2011}
{Dong}, R.,  \& {Draine}, B.~T. 2011, \apj, 727, 35

\bibitem[\protect\citeauthoryear{{Draine}}{{Draine}}{2011}]{Draine_2011a}
{Draine}, B.~T. 2011, {Physics of the Interstellar and Intergalactic Medium}
  (Princeton, NJ: Princeton University Press)

\bibitem[\protect\citeauthoryear{{Draine} \& {Li}}{{Draine} \&
  {Li}}{2007}]{Draine+Li_2007}
{Draine}, B.~T.,  \& {Li}, A. 2007, \apj, 657, 810

\bibitem[\protect\citeauthoryear{{Eisenstein} et~al.}{{Eisenstein}
  et~al.}{2011}]{Eisenstein+etal_2011}
{Eisenstein}, D.~J., {Weinberg}, D.~H., {Agol}, E., et~al. 2011, \aj, 142, 72

\bibitem[\protect\citeauthoryear{{Els{\"a}sser} \& {Haug}}{{Els{\"a}sser} \&
  {Haug}}{1960}]{Elsasser+Haug_1960}
{Els{\"a}sser}, H.,  \& {Haug}, U. 1960, \zap, 50, 121

\bibitem[\protect\citeauthoryear{{Elvey} \& {Roach}}{{Elvey} \&
  {Roach}}{1937}]{Elvey+Roach_1937}
{Elvey}, C.~T.,  \& {Roach}, F.~E. 1937, \apj, 85, 213

\bibitem[\protect\citeauthoryear{{Finkbeiner}}{{Finkbeiner}}{2003}]{Finkbeiner%
_2003}
{Finkbeiner}, D.~P. 2003, \apjs, 146, 407

\bibitem[\protect\citeauthoryear{{Finkbeiner}, {Davis}, \&
  {Schlegel}}{{Finkbeiner} et~al.}{1999}]{Finkbeiner+Davis+Schlegel_1999}
{Finkbeiner}, D.~P., {Davis}, M.,  \& {Schlegel}, D.~J. 1999, \apj, 524, 867

\bibitem[\protect\citeauthoryear{{Frieman} et~al.}{{Frieman}
  et~al.}{2008}]{Frieman+Bassett+Becker+etal_2008}
{Frieman}, J.~A., {Bassett}, B., {Becker}, A., et~al. 2008, \aj, 135, 338

\bibitem[\protect\citeauthoryear{{Gaustad} et~al.}{{Gaustad}
  et~al.}{2001}]{Gaustad+McCullough+Rosing+VanBuren_2001}
{Gaustad}, J.~E., {McCullough}, P.~R., {Rosing}, W.,  \& {Van Buren}, D. 2001,
  \pasp, 113, 1326

\bibitem[\protect\citeauthoryear{{Gilmore} \& {Reid}}{{Gilmore} \&
  {Reid}}{1983}]{Gilmore+Reid_1983}
{Gilmore}, G.,  \& {Reid}, N. 1983, \mnras, 202, 1025

\bibitem[\protect\citeauthoryear{{Gordon}, {Witt}, \& {Friedmann}}{{Gordon}
  et~al.}{1998}]{Gordon+Witt+Friedmann_1998}
{Gordon}, K.~D., {Witt}, A.~N.,  \& {Friedmann}, B.~C. 1998, \apj, 498, 522

\bibitem[\protect\citeauthoryear{{Henry}}{{Henry}}{1981}]{Henry_1981}
{Henry}, R.~C. 1981, \apjl, 244, L69

\bibitem[\protect\citeauthoryear{{Hern{\'a}ndez}, {Avila-Reese}, \&
  {Firmani}}{{Hern{\'a}ndez} et~al.}{2001}]{Hernandez+Avila-Reese+Firmani_2001}
{Hern{\'a}ndez}, X., {Avila-Reese}, V.,  \& {Firmani}, C. 2001, \mnras, 327,
  329

\bibitem[\protect\citeauthoryear{{Hudson} \& {Bell}}{{Hudson} \&
  {Bell}}{2005}]{Hudson+Bell_2005}
{Hudson}, C.~E.,  \& {Bell}, K.~L. 2005, \aap, 430, 725

\bibitem[\protect\citeauthoryear{{Hurwitz}, {Bowyer}, \& {Martin}}{{Hurwitz}
  et~al.}{1991}]{Hurwitz+Bowyer+Martin_1991}
{Hurwitz}, M., {Bowyer}, S.,  \& {Martin}, C. 1991, \apj, 372, 167

\bibitem[\protect\citeauthoryear{{Lillie} \& {Witt}}{{Lillie} \&
  {Witt}}{1976}]{Lillie+Witt_1976}
{Lillie}, C.~F.,  \& {Witt}, A.~N. 1976, \apj, 208, 64

\bibitem[\protect\citeauthoryear{{Madsen}, {Reynolds}, \& {Haffner}}{{Madsen}
  et~al.}{2006}]{Madsen+Reynolds+Haffner_2006}
{Madsen}, G.~J., {Reynolds}, R.~J.,  \& {Haffner}, L.~M. 2006, \apj, 652, 401

\bibitem[\protect\citeauthoryear{{Ma{\'{\i}}z Apell{\'a}niz}, {Alfaro}, \&
  {Sota}}{{Ma{\'{\i}}z Apell{\'a}niz}
  et~al.}{2008}]{Maiz-Apellaniz+Alfaro+Sota_2008}
{Ma{\'{\i}}z Apell{\'a}niz}, J., {Alfaro}, E.~J.,  \& {Sota}, A. 2008, ArXiv
  e-prints, 0804.2553

\bibitem[\protect\citeauthoryear{{Ma{\'{\i}}z-Apell{\'a}niz}
  et~al.}{{Ma{\'{\i}}z-Apell{\'a}niz}
  et~al.}{2004}]{Maiz-Apellaniz+Walborn+Galue+Wei_2004}
{Ma{\'{\i}}z-Apell{\'a}niz}, J., {Walborn}, N.~R., {Galu{\'e}}, H.~{\'A}.,  \&
  {Wei}, L.~H. 2004, \apjs, 151, 103

\bibitem[\protect\citeauthoryear{{Malhotra}}{{Malhotra}}{1995}]{Malhotra_1995}
{Malhotra}, S. 1995, \apj, 448, 138

\bibitem[\protect\citeauthoryear{{Martin}, {Hurwitz}, \& {Bowyer}}{{Martin}
  et~al.}{1990}]{Martin+Hurwitz+Bowyer_1990}
{Martin}, C., {Hurwitz}, M.,  \& {Bowyer}, S. 1990, \apj, 354, 220

\bibitem[\protect\citeauthoryear{{Mathis}, {Mezger}, \& {Panagia}}{{Mathis}
  et~al.}{1983}]{Mathis+Mezger+Panagia_1983}
{Mathis}, J.~S., {Mezger}, P.~G.,  \& {Panagia}, N. 1983, \aap, 128, 212

\bibitem[\protect\citeauthoryear{{Morgan}, {Nandy}, \& {Thompson}}{{Morgan}
  et~al.}{1978}]{Morgan+Nandy+Thompson_1978}
{Morgan}, D.~H., {Nandy}, K.,  \& {Thompson}, G.~L. 1978, \mnras, 185, 371

\bibitem[\protect\citeauthoryear{{Murthy} et~al.}{{Murthy}
  et~al.}{1990}]{Murthy+Henry+Feldman+Tennyson_1990}
{Murthy}, J., {Henry}, R.~C., {Feldman}, P.~D.,  \& {Tennyson}, P.~D. 1990,
  \aap, 231, 187

\bibitem[\protect\citeauthoryear{{Murthy}, {Henry}, \& {Holberg}}{{Murthy}
  et~al.}{1991}]{Murthy+Henry+Holberg_1991}
{Murthy}, J., {Henry}, R.~C.,  \& {Holberg}, J.~B. 1991, \apj, 383, 198

\bibitem[\protect\citeauthoryear{{Nakanishi} \& {Sofue}}{{Nakanishi} \&
  {Sofue}}{2003}]{Nakanishi+Sofue_2003}
{Nakanishi}, H.,  \& {Sofue}, Y. 2003, \pasj, 55, 191

\bibitem[\protect\citeauthoryear{{Neugebauer} et~al.}{{Neugebauer}
  et~al.}{1984}]{Neugebauer+Habing+vanDuinen+etal_1984}
{Neugebauer}, G., {Habing}, H.~J., {van Duinen}, R., et~al. 1984, \apjl, 278,
  L1

\bibitem[\protect\citeauthoryear{{Planck Collaboration} et~al.}{{Planck
  Collaboration} et~al.}{2011}]{Planck_2011_dust}
{Planck Collaboration}, {Ade}, P.~A.~R., {Aghanim}, N., et~al. 2011, ArXiv
  e-prints, 1101.2029

\bibitem[\protect\citeauthoryear{{Reynolds}, {Haffner}, \& {Madsen}}{{Reynolds}
  et~al.}{2002}]{Reynolds+Haffner+Madsen_2002}
{Reynolds}, R.~J., {Haffner}, L.~M.,  \& {Madsen}, G.~J. 2002, in \aspcf, Vol.
  282, Galaxies: the Third Dimension, ed. {M.~Rosada, L.~Binette, \& L.~Arias},
  31

\bibitem[\protect\citeauthoryear{{Sasseen} \& {Deharveng}}{{Sasseen} \&
  {Deharveng}}{1996}]{Sasseen+Deharveng_1996}
{Sasseen}, T.~P.,  \& {Deharveng}, J.-M. 1996, \apj, 469, 691

\bibitem[\protect\citeauthoryear{{Schlegel}, {Finkbeiner}, \&
  {Davis}}{{Schlegel} et~al.}{1998}]{Schlegel+Finkbeiner+Davis_1998}
{Schlegel}, D.~J., {Finkbeiner}, D.~P.,  \& {Davis}, M. 1998, \apj, 500, 525

\bibitem[\protect\citeauthoryear{{Seon} et~al.}{{Seon}
  et~al.}{2011}]{Seon+Edelstein+Korpela+etal_2010}
{Seon}, K.-I., {Edelstein}, J., {Korpela}, E., et~al. 2011, \apjs, 196, 15

\bibitem[\protect\citeauthoryear{{Stoughton} et~al.}{{Stoughton}
  et~al.}{2002}]{Stoughton+Lupton+Bernardi+etal_2002}
{Stoughton}, C., {Lupton}, R.~H., {Bernardi}, M., et~al. 2002, \aj, 123, 485

\bibitem[\protect\citeauthoryear{{Szomoru} \& {Guhathakurta}}{{Szomoru} \&
  {Guhathakurta}}{1998}]{Szomoru+Guhathakurta_1998}
{Szomoru}, A.,  \& {Guhathakurta}, P. 1998, \apjl, 494, L93

\bibitem[\protect\citeauthoryear{{Tayal} \& {Zatsarinny}}{{Tayal} \&
  {Zatsarinny}}{2010}]{Tayal+Zatsarinny_2010}
{Tayal}, S.~S.,  \& {Zatsarinny}, O. 2010, \apjs, 188, 32

\bibitem[\protect\citeauthoryear{{Weingartner} \& {Draine}}{{Weingartner} \&
  {Draine}}{2001}]{Weingartner+Draine_2001a}
{Weingartner}, J.~C.,  \& {Draine}, B.~T. 2001, \apj, 548, 296

\bibitem[\protect\citeauthoryear{{Witt} et~al.}{{Witt}
  et~al.}{2010}]{Witt+Gold+Barnes+etal_2010}
{Witt}, A.~N., {Gold}, B., {Barnes}, F.~S., III, et~al. 2010, \apj, 724, 1551

\bibitem[\protect\citeauthoryear{{Witt} et~al.}{{Witt}
  et~al.}{2008}]{Witt+Mandel+Sell+etal_2008}
{Witt}, A.~N., {Mandel}, S., {Sell}, P.~H., {Dixon}, T.,  \& {Vijh}, U.~P.
  2008, \apj, 679, 497

\bibitem[\protect\citeauthoryear{{Witt} \& {Vijh}}{{Witt} \&
  {Vijh}}{2004}]{Witt+Vijh_2004}
{Witt}, A.~N.,  \& {Vijh}, U.~P. 2004, in \aspcf\ 309, Astrophysics of Dust,
  ed. A.~N. {Witt}, G.~C. {Clayton}, \& B.~T. {Draine} (San Francisco, CA:
  ASP), 115

\bibitem[\protect\citeauthoryear{{Wolstencroft} \& {Rose}}{{Wolstencroft} \&
  {Rose}}{1966}]{Wolstencroft+Rose_1966}
{Wolstencroft}, R.~D.,  \& {Rose}, L.~J. 1966, \nat, 209, 388

\bibitem[\protect\citeauthoryear{{Wood} \& {Reynolds}}{{Wood} \&
  {Reynolds}}{1999}]{Wood+Reynolds_1999}
{Wood}, K.,  \& {Reynolds}, R.~J. 1999, \apj, 525, 799

\bibitem[\protect\citeauthoryear{{Yahata} et~al.}{{Yahata}
  et~al.}{2007}]{Yahata+Yonehara+Suto+etal_2007}
{Yahata}, K., {Yonehara}, A., {Suto}, Y., et~al. 2007, \pasj, 59, 205

\bibitem[\protect\citeauthoryear{{York} et~al.}{{York}
  et~al.}{2000}]{York+Adelman+Anderson+etal_2000}
{York}, D.~G., {Adelman}, J., {Anderson}, J.~E., Jr., et~al. 2000, \aj, 120,
  1579

\bibitem[\protect\citeauthoryear{{Zubko}, {Dwek}, \& {Arendt}}{{Zubko}
  et~al.}{2004}]{Zubko+Dwek+Arendt_2004}
{Zubko}, V., {Dwek}, E.,  \& {Arendt}, R.~G. 2004, \apjs, 152, 211

\bibitem[\protect\citeauthoryear{{Zvereva} et~al.}{{Zvereva}
  et~al.}{1982}]{Zvereva+Severnyi+Granitskii+etal_1982}
{Zvereva}, A.~M., {Severnyi}, A.~B., {Granitskii}, L.~V., et~al. 1982, \aap,
  116, 312

\end{thebibliography}

\end{document}